\documentclass[apj]{emulateapj}
\usepackage{graphics,graphicx,times,natbib,longtable,placeins}

\newcommand{\sersic}{{S\'{e}rsic }}

\def\mstar{${\rm M_{*}}$}
\def\msun{${\rm\,M_\odot}$}
\def\logm{{\rm log(M_{*}/M_\odot)}}

\def\fclumpy{$f_{clumpy}$}

\def\fluv{$f_{LUV}$}
 
\def\hst{{\it HST}}

\def\spose#1{\hbox to 0pt{#1\hss}}
\def\lta{\mathrel{\spose{\lower 3pt\hbox{$\mathchar"218$}}
     \raise 2.0pt\hbox{$\mathchar"13C$}}}

\shorttitle{Clump Properties at $0.5\leq z<3$}
\shortauthors{Guo et al.}

\begin{document}

\title{Clumpy Galaxies in CANDELS. II. Physical Properties of UV-bright Clumps
at $0.5\leq z<3$}

\author{Yicheng Guo\altaffilmark{1,2}}
\author{Marc Rafelski\altaffilmark{3}}
\author{Eric F. Bell\altaffilmark{4}}
\author{Christopher J. Conselice\altaffilmark{5}}
\author{Avishai Dekel\altaffilmark{6}}
\author{S. M. Faber\altaffilmark{1}}
\author{Mauro Giavalisco\altaffilmark{7}}
\author{Anton M. Koekemoer\altaffilmark{3}}
\author{David C. Koo\altaffilmark{1}}
\author{Yu Lu\altaffilmark{8}}
\author{Nir Mandelker\altaffilmark{9}}
\author{Joel R. Primack\altaffilmark{10}}
\author{Daniel Ceverino\altaffilmark{11}}
\author{Duilia F. de Mello\altaffilmark{12}}
\author{Henry C. Ferguson\altaffilmark{3}}
\author{Nimish Hathi\altaffilmark{3}}
\author{Dale Kocevski\altaffilmark{13}}
\author{Ray Lucas\altaffilmark{3}}
\author{Pablo G. P\'erez-Gonz\'alez\altaffilmark{14}}
\author{Swara Ravindranath\altaffilmark{3}}
\author{Emmaris Soto\altaffilmark{12}}
\author{Amber Straughn\altaffilmark{15}}
\author{Weichen Wang\altaffilmark{16}}
\altaffiltext{1}{UCO/Lick Observatory, Department of Astronomy and Astrophysics, University of California, Santa Cruz, CA, USA; {\it ycguo@ucolick.org}}
\altaffiltext{2}{Department of Physics and Astronomy, University of Missouri, Columbia, MO, USA; {\it guoyic@missouri.edu}}
\altaffiltext{3}{Space Telescope Science Institute, Baltimore, MD, USA}
\altaffiltext{4}{Department of Astronomy, University of Michigan, Ann Arbor, MI, USA}
\altaffiltext{5}{School of Physics and Astronomy, University of Nottingham, University Park, Nottingham NG7 2RD, UK}
\altaffiltext{6}{Center for Astrophysics and Planetary Science, Racah Institute of Physics, The Hebrew University, Jerusalem, Israel}
\altaffiltext{7}{Department of Astronomy, University of Massachusetts, Amherst, MA, USA}
\altaffiltext{8}{Observatories, Carnegie Institution for Science, Pasadena, CA, USA}
\altaffiltext{9}{Department of Astronomy, Yale University, New Haven, CT, USA}
\altaffiltext{10}{Department of Physics, University of California, Santa Cruz, CA, USA}
\altaffiltext{11}{Institut f{\"u}r Theoretische Astrophysik, Zentrum f{\"u}r Astronomie der Universit{\"a}t Heidelberg, Albert-Ueberle-Str. 2, D-69120 Heidelberg, Germany}
\altaffiltext{12}{Physics Department, The Catholic University of America, Washington DC, USA}
\altaffiltext{13}{Physics and Astronomy Department, Colby College, Waterville, ME, USA}
\altaffiltext{14}{Departamento de Astrof\'{\i}sica, Facultad de CC.  F\'{\i}sicas, Universidad Complutense de Madrid, E-28040 Madrid, Spain}
\altaffiltext{15}{Astrophysics Science Division, Goddard Space Flight Center, Code 665, Greenbelt, MD 20771, USA}
\altaffiltext{16}{Department of Physics \& Astronomy, Johns Hopkins University, Baltimore, MD, USA}


\begin{abstract} 

Studying giant star-forming clumps in distant galaxies is important to
understand galaxy formation and evolution. At present, however, observers and
theorists have not reached a consensus on whether the observed ``clumps'' in
distant galaxies are the same phenomenon that is seen in simulations. In this
paper, as a step to establish a benchmark of direct comparisons between
observations and theories, we publish a sample of clumps constructed to
represent the commonly observed ``clumps'' in the literature. This sample
contains 3193 clumps detected from 1270 galaxies at $0.5 \leq z < 3.0$. The
clumps are detected from rest-frame UV images, as described in our previous
paper. Their physical properties, e.g., rest-frame color, stellar mass
(\mstar), star formation rate (SFR), age, and dust extinction, are measured by
fitting the spectral energy distribution (SED) to synthetic stellar population
models. We carefully test the procedures of measuring clump properties,
especially the method of subtracting background fluxes from the diffuse
component of galaxies. With our fiducial background subtraction, we find a
radial clump U-V color variation, where clumps close to galactic centers are
redder than those in outskirts. The slope of the color gradient (clump color as
a function of their galactocentric distance scaled by the semi-major axis of
galaxies) changes with redshift and \mstar\ of the host galaxies: at a fixed
\mstar, the slope becomes steeper toward low redshift; and at a fixed redshift,
it becomes slightly steeper with \mstar. Based on our SED-fitting, this
observed color gradient can be explained by a combination of a negative age
gradient, a negative E(B-V) gradient, and a positive specific star formation
rate gradient of the clumps.  We also find that the color gradients of clumps
are steeper than those of intra-clump regions. Correspondingly, the radial
gradients of the derived physical properties of clumps are different from those
of the diffuse component or intra-clump regions. 

\end{abstract}

\section{Introduction}
\label{intro}

\subsection{Overview: Clumps and Their Formation and Evolution}
\label{intro:overview}

To understand how the morphology and structure of galaxies evolve over cosmic
time requires knowledge of not only integrated galaxy properties, but also
sub-structures of galaxies. Current facilities enable us to resolve distant
galaxies and study their spatially resolved physical properties, including (I)
{\it sub-structures}
\citep[e.g.,][]{elmegreen05,elmegreen07,elmegreen09a,elmegreen09b,genzel08,genzel11,fs11b,ycguo12clump,ycguo15fclumpy,wuyts12,tadaki14,shibuya16,soto17},
(II) {\it color variation}
\citep[e.g.,][]{menanteau04,mcgrath08,tortora10,gargiulo11,gargiulo12,ycguo11peg,szomoru11,boada15,tacchella15b,chan16,liufs16},
(III) {\it star formation variation}
\citep[e.g.,][]{wuyts13,hemmati14,hemmati15,tacchella15a,barro16,mieda16,nelson16a,nelson16b},
and (IV) {\it mass distribution and central concentration}
\citep[e.g.,][]{saracco12,szomoru13,lang14,vandokkum14,barro15,mosleh17}.

A common and important sub-structure of distant star-forming galaxies is giant
off-center star-forming clumps. These clumps are seen in deep and
high-resolution rest-frame UV and optical images
\citep[e.g.,][]{conselice04,elmegreen05,elmegreen07,elmegreen09a,fs11b,ycguo12clump,ycguo15fclumpy,wuyts12,murata14,tadaki14,shibuya16,soto17}.
They are also detected in high-resolution emission line maps of H$\alpha$
\citep[e.g.,][]{genzel08,genzel11,livermore12,livermore15,wisnioski11,mieda16,fisher17b}
and CO \citep[e.g.,][]{jones10,swinbank10,dz16}. The clumps appear to be much
larger, brighter, and more massive than local star-forming regions. Their
typical stellar mass (\mstar) is ${\rm 10^{7}-10^{9} M_\odot}$
\citep[e.g.,][]{elmegreen07,ycguo12clump,soto17}. Their actual sizes are
uncertain due to the resolution limit of current observations, ranging from
$\sim$1 kpc \citep[e.g.,][]{elmegreen07,fs11b} to a few hundred pc
\citep[e.g.,][]{livermore12}. The clumps resemble mini-starbursts in their
galaxies \citep[e.g.,][]{bournaud15,zanella15} and have specific star formation
rates (sSFR) higher than their surrounding areas by a factor of several,
evident by their blue UV--optical colors or enhanced H$\alpha$ surface
brightness \citep[e.g.,][]{ycguo12clump,wuyts12,wuyts13,hemmati14,mieda16}.

The formation and evolution of clumps provide important tests of our knowledge
of star formation, feedback, and galactic structure formation. Clumps are
thought to form through gravitational instability in gas-rich turbulent disks
\citep[e.g.,][]{noguchi99,immeli04a,immeli04b,bournaud07,bournaud09a,elmegreen08,dekel09,ceverino10,ceverino12,dekel14,inoue16}.
This view is supported by some observations, especially for massive clumpy
galaxies
\citep[e.g.,][]{elmegreen07,bournaud08,genzel08,genzel11,ycguo12clump,ycguo15fclumpy,hg16,mieda16,fisher17b}.
The kinematic signatures of the clumpy galaxies, however, can also have an
ex-situ origin, such as gas-rich mergers
\citep[e.g.,][]{hopkins13}, which also has some supporting observations
\citep[e.g.,][]{puech09,puech10,wuyts14,ycguo15fclumpy,straughn15,ribeiro16}.

The evolution of clumps is under intense debate. Some models predict that they
would migrate toward the gravitational centers of their host galaxies, due to
clump--clump and clump--disk interactions and dynamical friction, and
eventually coalesce into a young bulge as a progenitor of today's bulges
\citep[e.g.,][]{bournaud07,elmegreen08,ceverino10,mandelker14,bournaud14a}.
Observational evidence of this scenario is the age (or color) variation of
clumps with galactocentric distance (clump age gradient). Some studies, e.g.,
\citet{fs11b}, \citet{ycguo12clump}, \citet{shibuya16}, and \citet{soto17}
found that clumps close to galactic centers are older than those in the
outskirts, broadly consistent with the prediction of the inward migration
scenario. In this scenario, clumps need to survive longer than $\sim$150 Myr to
be able to travel to galactic centers.

On the other hand, some models predict a short life time of clumps ($\lesssim
50$ Myr) because of quick disruption of clumps by either tidal forces or
stellar feedback
\citep[e.g.,][]{murray10,genel12,hopkins12clump,buck16,oklopcic17}.  The
disrupted stars from clumps would contribute to the formation of thick disks
\citep[e.g.,][]{bassett14,inoue14,struck17}. Clumps have high sSFR and
therefore strong star formation feedback for their \mstar\
\citep[e.g.,][]{genzel08,newman12clump}, which enables the quick disruption,
although clumps with typical star formation efficiency of a few percent per
free-fall time are not disrupted \citep{krumholz10}.
Although the observed stellar age gradient and the older-than-100 Myr age of
some clumps seem contradictory to the quick disruption scenario, a few
simulations, e.g., \citet{buck16} and \citet{oklopcic17}, argue that these
observations can be interpreted as clump regions being contaminated by older
disk stars, and therefore may not be an indicator of the long lifetime and
inward migration of clumps.

Understanding the evolution of clumps is important, because it reveals whether
clumps are a major contributor to bulge formation, given their prevalence at
high redshifts \citep[e.g.,][]{elmegreen07,tadaki14,ycguo15fclumpy,shibuya16}.
Clumps may also significantly contribute to the growth of supermassive black
holes and AGN \citep[e.g.,][]{bournaud11,gabor13}, which has both supporting
\citep{bournaud12,ycguo12clump} and contradicting observational evidence
\citep{trump14}.

Whether or not clumps are a major contributor to bulge formation, they are
important to further understand star formation feedback, because they provide a
sensitive diagnostic of feedback models on sub-galactic scales. \citet{moody14}
showed that adding radiation pressure to a feedback recipe significantly
reduces the number of intermediate-mass clumps (\mstar$\lesssim 10^{8}$\msun)
in their Adaptive Refinement Tree (ART) simulations
\citep{kravtsov97,kravtsov03,ceverino09}, compared to having only supernova
feedback. \citet{mandelker17} further studied the dependence of clump
properties and evolution on feedback recipes in detail. They found that
including radiation pressure would increase the baryonic surface density and
baryonic mass thresholds for clumps that are long-lived and not being disrupted
in a few free-fall times. In the FIRE simulation \citep{hopkins14fire}, where
the feedback recipe reduces star formation more than in the ART simulations,
clumps are reported to all have short lifetimes $\lesssim$50 Myr
\citep{oklopcic17}. Therefore, the properties of clumps, e.g., number, mass,
star formation rate (SFR), and age, are important to test the validity of
feedback models.

\subsection{A Challenge of Clump Studies and the Motivation of This Paper}
\label{intro:challenge}

The resolution and sensitivity of current facilities pose a significant
challenge when comparing observations to models, and even observations to each
other.
Even with {\it HST}, clumps at high redshifts can only be marginally resolved
or may even be unresolved. This limitation raises the question whether an
observed clump is actually a single object or blending of a few nearby smaller
clumps. For example, \citet{tamburello15} argued that many of the giant clumps
(with \mstar$>10^8$\msun) identified in observations are not due to in-situ
formation, but are the result of blending of smaller structures due to the low
resolution of observations. Similarly, \citet{dz17} claimed that the clump
masses observed in non-lensed galaxies with a limited spatial resolution of
$\sim$1 kpc are artificially increased due to the clustering of clumps of
smaller mass. They also stated that the sensitivity threshold used for the
clump selection strongly biases against clumps at the low-mass end. Similarly,
\citet{fisher17a} discussed the effects of clump clustering on kpc
scale measurements of clumps. They inferred that the clustering systematically
increases the apparent size and SFR of clumps in 1 kpc resolution maps, and
decreases the measured SFR surface density of clumps by as much as a factor of
20. 

To address the issue of observational effects, in \citet[][hereafter Paper
I]{ycguo15fclumpy}, we proposed a physical definition that UV-bright clumps are
off-center discrete star-forming regions that individually contribute more than
8\% of the rest-frame UV light of their galaxies. This definition is derived
through comparison with redshifted and size-matched nearby spiral galaxies.
Clumps defined this way are significantly brighter than the HII regions of
nearby large spiral galaxies, either individually or blended, when physical
spatial resolution and cosmological dimming are considered. This objective and
physical definition enables a consistent study of clumps at different
redshifts, 
complementing clump studies of using the {\it appearance} of galaxies by either
visual inspection
\citep[e.g.,][]{cowie95,vandenbergh96,elmegreen04,elmegreen05,elmegreen07} or
other automated algorithm
\citep[e.g.,][]{conselice03,conselice04,fs11b,ycguo12clump,wuyts12,murata14}.

Under this definition, in Paper I, we measured the fraction of star-forming
galaxies that have at least one off-center clump ($f_{clumpy}$). The redshift
evolution of $f_{clumpy}$ changes with \mstar\ of the host galaxies. Low-mass
($\logm < 9.8$) galaxies keep an almost constant $f_{clumpy}$ of $\sim$60\%
from $z\sim3$ to $z\sim0.5$. Intermediate-mass ($9.8 \leq \logm < 10.6$) and
massive ($\logm \geq 10.6$) galaxies drop their $f_{clumpy}$ from 55\% at
$z\sim3$ to 40\% and 15\%, respectively, at $z\sim0.5$. We found that (1) the
trend of disk stabilization predicted by violent disk instability matches the
\fclumpy\ trend of massive galaxies; (2) minor mergers are a viable explanation
of the \fclumpy\ trend of intermediate-mass galaxies at $z<1.5$, given a
realistic observability timescale; and (3) major mergers are unlikely
responsible for the \fclumpy\ trend in all masses at $z<1.5$. 

\begin{figure*}[htbp]
\center{\hspace*{-0.5cm}\includegraphics[scale=0.3, angle=0]{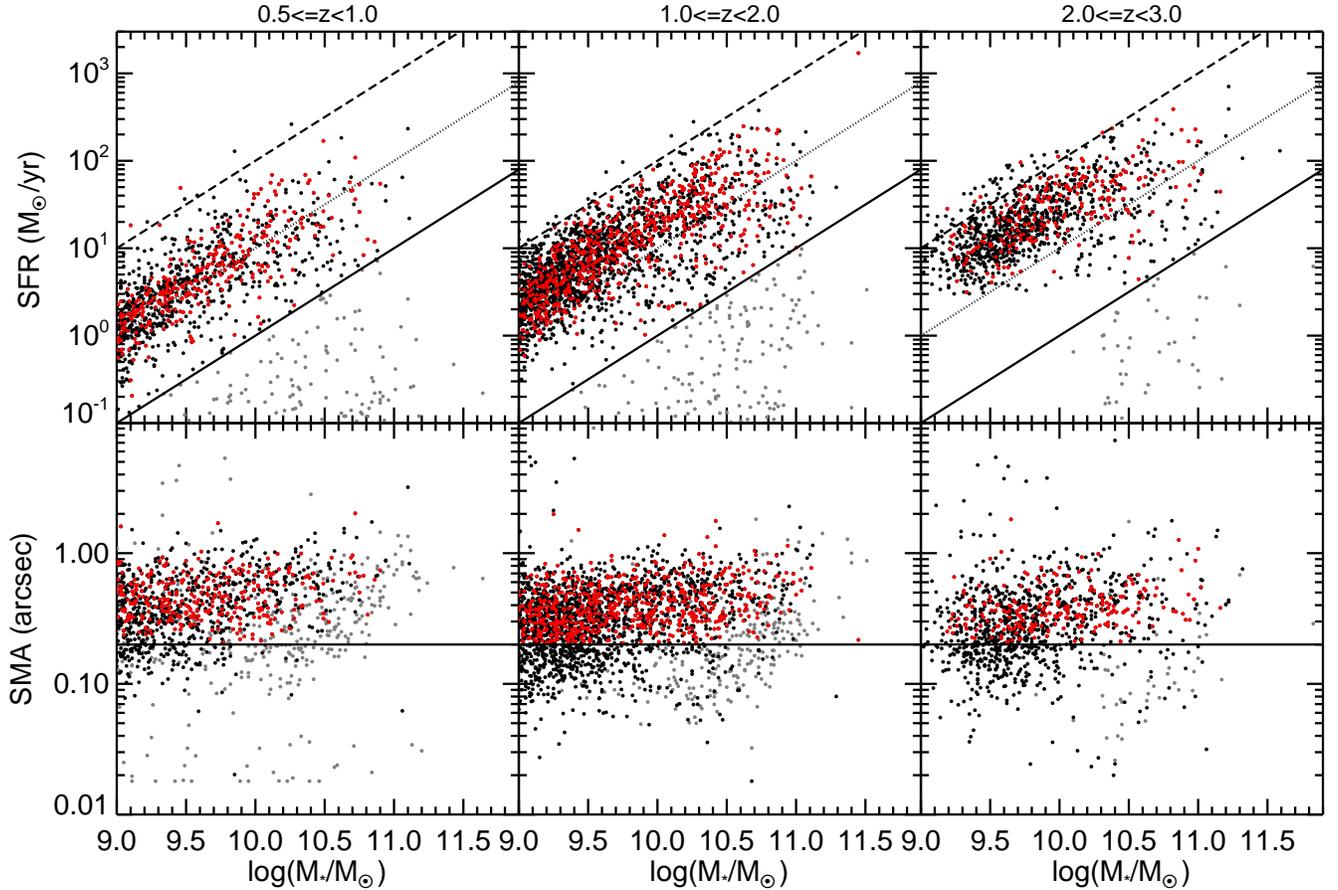}}

\caption[]{Clumpy galaxy sample split into three redshift bins. Galaxies in the CANDELS/GOODS-S with
$H_{F160W} < 24.5$ AB are plotted in the SFR--\mstar\ and semi-major axis
(SMA)--\mstar\ diagrams. Galaxies with sSFR$> 0.1 Gyr^{-1}$ are black, while
those with sSFR$\leq 0.1 Gyr^{-1}$ are gray. 
Galaxies with at least one detected off-center clump (with $f_{LUV} \equiv
L_{clump}^{UV} / L_{galaxy}^{UV} \geq 3\%$) are red. The red points all have
axial ratio $q > 0.5$.
Black solid, dotted, and dashed lines in the upper panels show the relations of
sSFR=0.1, 1, and 10 $Gyr^{-1}$. Black horizontal lines in the lower panels show
our size cut of 0\farcs2. 

\label{fig:sample}}
\vspace{-0.2cm}
\end{figure*}

This paper is the second of a series aiming to understand the observational
effects of clump studies and hence constructing a direct and unbiased
comparison between observation and theory. We measure the physical properties
of the clumps detected in Paper I and provide our clump catalog to the
community. The goal of this paper is not to use our measurements to test
models. Rather, it is to present a sample which, to the best of our knowledge,
represents the observed ``clumps'' in the literature. At present, observers and
theorists have not reached a consensus on whether the observed ``clumps'' are
the same phenomenon that is seen in simulations. A critical step of reaching
the consensus is to understand the physical properties of the observed clump
over wide redshift and mass ranges. Public catalogs containing detailed
information of clumps, however, are still insufficient (for example, see the
compilation of \citet{dz17}). The catalog present in this paper is our
contribution to establish a benchmark of direct comparisons, providing a
dataset to allow (1) theorists to understand the observed ``clumps'' in a wide
range of redshifts and galaxy \mstar\ and (2) observers to examine the
observational effects for a large sample of clumps. 

In this paper, we briefly summarize the galaxy sample and clump sample in
Section \ref{sample}. In Section \ref{photometry}, we describe the measurement
of multi-band photometry of individual clumps and test its accuracy. We
particularly test the effects of different methods of subtracting the
surrounding background of clumps, because the contamination by disk stars is a
major uncertainty when interpreting the observed properties. In Section
\ref{measureprop}, we describe the measurement of the stellar population of
clumps. We also present a few sanity checks on the accuracy of the measurement.
In Section \ref{observedprop}, we show a few examples of the measured physical
properties, which we think may be of interest to most readers. In Section
\ref{catalog}, we briefly introduce the clump catalog and a few cautions for
using it.

Throughout the paper, we adopt a flat ${\rm \Lambda CDM}$ cosmology with
$\Omega_m=0.3$, $\Omega_{\Lambda}=0.7$ and use the Hubble constant in terms of
$h\equiv H_0/100 {\rm km~s^{-1}~Mpc^{-1}} = 0.70$. All magnitudes in the paper
are in AB scale \citep{oke74} unless otherwise noted. We use a Chabrier initial
mass function  \citep[IMF][]{chabrier03}. 

\section{Data and Sample Selection}
\label{sample}

\subsection{Galaxy Sample}
\label{sample:galaxy}

The galaxies used in this paper are from the CANDELS/GOODS-S sample presented
in Paper I, which is based on the CANDELS survey
\citep{candelsoverview,candelshst}. We do not include the CANDELS/UDS sample of
Paper I in this paper, because UDS only has four {\it HST} bands, not enough
for carrying out spatially-resolved SED-fitting for individual clumps. While
referring readers to Paper I for details, we briefly summarize key selection
criteria here.

The multi-band photometry catalog of GOODS-S is described by
\citet{ycguo13goodss}. The photometric redshift (photo-z) was measured by the
method of \citet{dahlen13}. \mstar\ and SFR are measured through SED-fitting.
CANDELS has generated a unified \mstar\ catalog \citep{mobasher15,santini15},
where each galaxy is fit by 12 SED-fitting codes with different combinations of
synthetic stellar population models, star formation histories, fitting methods,
etc. For each galaxy, we quote the median of the best-fit \mstar\ of the 12
SED-fitting codes as its \mstar. We also use the median SFR of the 12
SED-fitting codes as our SFR measurement.

Star-forming galaxies are selected to have \mstar${\rm > 10^9 M_\odot}$,
sSFR${\rm >10^{-1} Gyr^{-1}}$, and $0.5\leq z<3$. We also use an apparent
magnitude cut of $H_{F160W}<24.5$ AB to ensure a reliable morphology and size
measurements of the galaxies. We only use galaxies whose effective radii along
the galaxy semi-major axis (SMA) is larger than 0\farcs2, because clumps cannot
be resolved in smaller galaxies. To minimize the effect of dust extinction and
clump blending, we only use galaxies with axial ratio $q>0.5$.

After the above selection criteria, and further excluding galaxies that are not
covered by the ACS images, the sample used for detecting clumps consists of
1655 galaxies (some of them may not contain clumps). As a comparison, the
CANDELS/UDS sample in Paper I contains 1584 galaxies.

\subsection{Clump Sample}
\label{sample:clump}

Clump detection is detailed in Paper I. Briefly, clumps are detected in
rest-frame Near-UV ($\sim 2800 \AA$), i.e., ACS F435W at $0.5\leq z<1.0$, F606W
at $1.0\leq z<2.0$, and F775W at $2.0\leq z<3.0$. First, the detection image is
smoothed. The smoothed image is then subtracted from the original image to make
a contrast image. After low-S/N pixels are masked out, only off-center
``blobs'' are detected from the filtered image as regions with at least five
contiguous pixels. We choose rest-frame Near-UV because it is observed by {\it
HST} filters for the whole GOODS-S field across the redshift range of
$0.5\leq z<3.0$. Some clumps are brighter and more prominent in rest-frame
Far-UV ($\sim 1500 \AA$) as found by \citet{soto17}, which makes Far-UV also an
efficient band to detect active star-forming clumps. However, deep Far-UV
observations are only available for about one-third area of GOODS-S for
galaxies at $0.5\leq z<1.5$. To increases our sample size, we use the Near-UV
detection in this paper. Future large and deep Far-UV surveys are needed to
promote studies of Far-UV clumps.

In Paper I, we only refer to those ``blobs'' which contribute at least 8\% of
the total UV luminosity of their galaxies as clumps (see Section \ref{intro}
for the relevant discussion). This objective and physical definition is
necessary when considering clumps as a distinct feature from normal nearby HII
regions and studying the redshift evolution of clumps and clumpy galaxies (as
in Paper I). This definition, however, has a few caveats. First, this
definition only selects very UV-bright star-forming regions with $f_{LUV}
\equiv L_{clump}^{UV} / L_{galaxy}^{UV} > 8\%$ and identifies all fainter
star-forming regions as non-clumps. This bisection of star-forming regions
implies an abrupt change of the properties of star formation regions.
Theoretical models, however, predict a wide and continuous distribution
of clumps properties \citep[e.g.,][]{moody14,mandelker17}. Second, the number
of clumps is predicted to increase toward low luminosity \citep{mandelker17}.
Excluding fainter star-forming regions results in an incomplete and biased
clump sample. Third, the threshold of 8\% is determined based on
redshifting only one local galaxy (M101) in Paper I and may not fully
represent all clumps. In this paper, we tend to be more inclusive and include
blobs down to $f_{LUV} = 3\%$. This inclusion enlarges our clump sample, but it
may also include regions that are similar to normal nearby HII regions. We
leave to readers to decide a preferred $f_{LUV}$ threshold.  Overall, the
sample has 1547 clumps with $f_{LUV} \geq 8\%$, 854 clumps with $5\% \leq
f_{LUV} < 8\%$, and 792 with $3\% \leq f_{LUV} < 5\%$. In total, 3193 clumps
are detected from 1270 galaxies. Figure \ref{fig:sample} shows these clumpy
galaxies in the SFR--\mstar\ and SMA--\mstar\ diagrams.

\begin{figure}[htbp] \center{
\includegraphics[scale=0.45, angle=0]{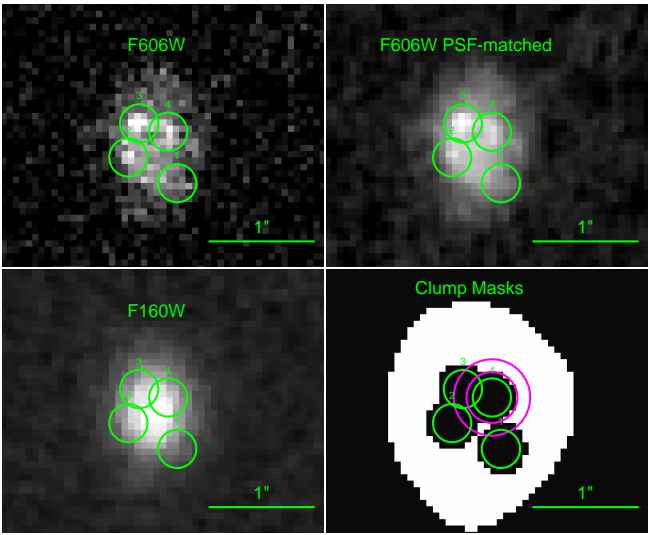}}

\caption[]{Illustration of diffuse light subtraction. A galaxy (Galaxy ID =
25508 in the catalog) is shown in ACS F606W used to detect clumps (top left),
smoothed ACS F606W to match the resolution of WFC3 F160W (top right), WFC3
F160W (bottom left), and a mask image (bottom right). Four clumps are detected
in this galaxy as shown by the green circles with radius of 0\farcs18 (3
pixels) in each panel. In the mask image (bottom right), the area within
0\farcs24 (4 pixels) of the center of each clump is masked out (i.e., black
pixels in the panel). These pixels are not used in calculating the diffuse
background. The pixels outside the galaxy are also masked out, because they are
out of the SExtractor segmentation map of the galaxy. For one clump (Clump ID =
4 in the catalog), we show the annulus (magenta circles) used as our fiducial
method (bgsub\_v4 in Table \ref{tb:bkg}) to measure the surface brightness of
the diffuse background (or intra-clump regions). The annulus (between two
magenta circles) has the inner and outer radii of 0\farcs24 and 0\farcs36. Only
the white pixels (i.e., those not masked out due to clump locations) between
the two magenta circles are used to calculate the surface brightness of the
diffuse background.

\label{fig:bkg}}
\vspace{-0.2cm}
\end{figure}

\begin{table*}[htbp]
\center{
\caption{Number of Clumps in the Sample
\label{tb:clumpstat}}
\hspace*{-0.5cm}
\begin{tabular}{|c|c|c|c|}
\hline\hline
       & ${\rm log(M^{gal}_*/M_\odot)} < 9.8$\footnote{$M^{gal}_*$ means \mstar\ of host galaxies.}$^,$\footnote{In each ($M^{gal}_*$, $z$) bin, the three numbers in the third row are the numbers of clumps in three $f_{LUV}$ bins: 0.03--0.05, 0.05--0.08, and $>0.08$.} & $9.8 \leq {\rm log(M^{gal}_*/M_\odot)} < 10.6$ & ${\rm log(M^{gal}_*/M_\odot)} \geq 10.6$ \\
\hline
$0.5\leq z < 1.0$ & 230 galaxies & 113 galaxies & 16 galaxies \\ 
                  & 654 clumps & 304 clumps & 38 clumps \\ 
                  & 171/198/285 & 107/83/114 & 13/16/9 \\ 
\hline
$1.0\leq z < 2.0$ & 363 galaxies & 239 galaxies & 67 galaxies \\ 
                  & 947 clumps & 658 clumps & 140 clumps \\ 
                  & 182/273/492 & 218/173/267 & 60/32/48 \\ 
\hline
$2.0\leq z < 3.0$ & 93 galaxies & 127 galaxies & 22 galaxies \\ 
                  & 181 clumps & 230 clumps & 41 clumps \\ 
                  & 10/25/146 & 26/45/159 & 5/9/27 \\ 
\hline
\end{tabular}
} \\
\end{table*}

\subsection{Selection Effects}
\label{sample:bias}

To detect clumps in rest-frame NUV, we use different HST bands at different
redshifts: F435W ($B$) at $0.5\leq z < 1$; F606W ($V$) at $1\leq z < 2$; and
F775W ($i$) at $2\leq z < 3$. The clump detection is affected by the
sensitivities of the three bands. 
Paper I shows that the clump magnitude of 50\% completeness of our detection is
about 28.5 AB, 28.5 AB, and 27.8 AB in $B$, $V$, and $i$, respectively. As a
result, fainter clumps are harder to detect at higher redshifts.

A more relevant way to evaluate the incompleteness is calculating the
incompleteness as a function of \fluv\ rather than a function of the clump
apparent magnitude. Our clump definition is based on \fluv\ because (1) our
clump finder detects clumps from a contrast image, showing how bright the
clumps are relative to their host galaxies and (2) using a relative ratio of an
intrinsic parameter enables direct comparisons between observations with
different sensitivities. 
Paper I shows that at $0.5\leq z < 1.0$, $1.0\leq z < 2.0$, and $2.0 \leq z <
3.0$, the 50\% completeness occurs at \fluv=0.03, 0.06, and 0.10, respectively.
A large fraction of clumps with low \fluv\ is missed in our clump detection at
high redshift. For example, we might miss 90\% of clumps with \fluv=0.03 at
$z>2$. This effect raises an important caution of using our clump catalog: it
is a ``representative'' rather than a ``complete'' catalog. Incompleteness
needs to be taken into account when deriving measurements that require a
complete sample, e.g., stellar mass function of clumps.

Specifically, our relative definition of clumps, namely, selecting clumps based
on \fluv\ rather than on their absolute luminosity, would introduce two biases.
First, a redshift-dependent bias is introduced to clump measurement. At high
redshifts, only high \fluv\ clumps are detected. Therefore, the results are
dominated by high \fluv\ clumps. Moreover, high \fluv\ clumps at high
redshifts are intrinsically much brighter or more luminous than the
corresponding (i.e. same \fluv) clumps at lower redshift. In contrast, at low
redshifts, the results are contributed by all \fluv\ clumps. To help readers to
evaluate this effect, we list the number of clumps with different \fluv\ at
different ($z$, \mstar) bins in Table. \ref{tb:clumpstat}. 

The second bias is introduced by the total UV luminosity of galaxies with
different \mstar. Massive star-forming galaxies have higher UV luminosity than
lower-mass galaxies. Clumps detected from massive galaxies are therefore
intrinsically UV brighter than those from lower-mass galaxies. In other words,
faint clumps can only be selected from low UV luminosity galaxies; and from UV
bright galaxies, only UV bright clumps can be selected. This bias would mainly
affect comparisons of the distributions of clump properties (e.g., SFR
distribution) between galaxies with different \mstar. Both biases are
introduced because for a given \fluv\ class, the absolute clump UV luminosity
may vary considerably among the different redshift and/or galaxy \mstar\
ranges. The robustness of our results against both biases can be tested by
calculating the results using clumps with different \fluv\ separately (see
Section \ref{prop:clgrad_color}).

\section{Multi-band Clump Photometry}
\label{photometry}

For each clump, we measure its multi-band photometry from {\it HST} images in
the bands of F435W, F606W, F775W, F814W, F850LP, F105W, F125W,
F140W\footnote{This band is taken by 3D-HST \citep{brammer123dhst}.}, and
F160W. {\it HST} images have the capability to resolve at the kpc scale at
$0.5\leq z<3.0$. The images in all bands are PSF-matched to the resolution of
F160W (FWHM=0\farcs17) by using the IRAF/PSFMATCH package. Details of the
PSF-matching method are described in \citet{ycguo11peg,ycguo13goodss}. As shown
in Figure 5 of \citet{ycguo13goodss}, for an aperture of 3 pixels, the accuracy
of PSF-matching is within 5\%.

For each clump, we use an aperture of 0\farcs18 (3 pixels) to measure its flux
in each band. We then multiply all the fluxes by a factor of 1.5 for the
aperture correction. This factor is derived through the curve-of-growth of
F160W PSF (see the middle panel of Figure 5 of \citet{ycguo13goodss}). This
method assumes that each clump is an unresolved source, which is reasonable for
the {\it HST} resolution at F160W, because the FWHM of F160W (0\farcs17) is
corresponding to $\sim$1 kpc (and $\sim$1.4 kpc) at $z=0.5$ (and at
$1.5\lesssim z \lesssim 3.0$). The assumption of an unresolved source makes it
easy to calculate the clump light that is out of our fixed aperture.

\subsection{Diffuse Background Subtraction}
\label{photometry:bkg}

A challenge is how to subtract the light from the underlying diffuse component
(or intra-clump regions) of the galaxies.
Clumps are believed to be ``embedded'' in a diffuse background. Therefore, in a
clump location, the observed light is contributed by both clump stars and
background stars.  Separating the two contributions is scientifically critical.
Some numerical simulations \citep[e.g.,][]{buck16,oklopcic17} found a radial
variation of clump ages: older for inner (small galactocentric distance) clumps
and younger for outer (large galactocentric distance) clumps, which seems to
support the inward migration scenario (see discussion in Section \ref{intro}).
These simulations, however, found no clump migration, and they argued that the
clump age gradient in their simulations is the result of contamination of disk
(i.e., background) stars. To make a direct comparison with models, background
subtraction is important to eliminate the contamination.

Observationally, subtracting the diffuse background is important for measuring
clump photometry. After PSF-matching to the F160W resolution, some clumps may
disappear due to the reduced contrast between clumps and the background.
Moreover, some intrinsically blue clumps are almost ``invisible'' in red bands
(e.g., F160W) again due to small clump--background contrasts (e.g., see clumps
in Figure \ref{fig:bkg} and discussion in \citet{soto17}).  For these clumps,
we fix their centers as those detected in the non-PSF-matched Near-UV bands to
measure the aperture photometry. The accuracy of the photometry significantly
relies on the precision of the diffuse background subtraction, because the
photometry is in fact the measurement of the ``excess'' flux above the diffuse
background flux. 

To subtract the diffuse background light, in each band, we first mask out the
locations of all detected clumps with circular masks. Then, for each clump, we
measure the background flux from an annulus around the clump to calculate the
average local background surface brightness (i.e., flux per pixel) in the clump
vicinity. We then subtract the background flux within the clump aperture of
0\farcs18 (i.e., the local background surface brightness times the clump area).
The subtracted background for the clump is contributed by two sources: the
local diffuse background and the PSF wings of nearby clumps and the analyzed
clump itself. Figure \ref{fig:bkg} illustrates the method.

Two parameters control this method: (1) the size (radius) of the clump mask and
(2) the size of the background annulus. We try different combinations of them
(see Table \ref{tb:bkg}). For an aggressive subtraction (e.g., bgsub\_v6), the
clump mask size is small, leaving more pixels as background (or intra-clump)
pixels, and the background annulus starts right next to the clump region. In
contrast, a conservative subtraction (e.g., bgsub\_v1) masks more pixels as
clump regions and measures the background far away from the clump regions. As
shown in Table \ref{tb:bkg}, the difference between aggressive subtraction
(bgsub\_v6) and conservative subtraction (bgsub\_v1) is about a factor of 1.9
in clumps' F160W flux. If no background is subtracted (bgsub\_v0), the F160W
flux of clumps is about a factor of 3.2 higher than that of bgsub\_v6. Here we
use F160W flux to show the effect of background subtraction, because this issue
is most significant in F160W due to the faint clump fluxes in this band.

We choose method bgsub\_v4 as our fiducial one based on a test of fake clumpy
galaxies. In this test, we use the WFC3 F160W PSF as fake clumps and insert
four fake clumps into a constant diffuse background. For each fake galaxy, we
choose one clump as the target clump and normalize its flux to unity. The
fluxes of other fake clumps are randomly drawn from the range of 0.1--10
relative to the flux of the target clump. The separation between the fake
clumps is drawn from the observed distribution of clump--clump distances. 
The surface brightness of the constant background ranges from 0.01\% to 10\% of
the peak surface brightness of the target fake clump. 

We apply all subtraction methods to the fake galaxies and test which one
recovers the input flux of the target clump the best. The most affecting
parameter in this test is the flux ratio between the target fake clump and its
closest neighbor. On average, method bgsub\_v4 recovers the target flux the
best over the range of the clump flux ratio. It recovers $\gtrsim$95\% of the
target flux when the neighbor clump is not brighter than the target one by a
factor of two. It overestimates the target flux by a factor of 1.1 when the
neighbor clump is five times brighter. 

Background subtraction, however, is more complicated than the above test.
Technically, a few issues are not covered by the test. First, some clumps may
be more extended then WFC3 PSFs. Second, the diffuse background is not constant
in the galaxy-size scale. 
Third and more importantly, different clumps have different distances and flux
ratios to their neighbor clumps, which means the best background subtraction
configuration may vary from clump to clump. 

Physically, whether or not the background should be subtracted depends on the
scientific goals. For example, if the goal is to study the clump properties
(e.g., mass, SFR, etc.), the background needs to be subtracted. If, however,
the goal is to study the dynamics of clump regions, the background should be
kept. To enable as many topics as possible, we release the clump catalogs with
all the background subtraction methods (including no subtraction), allowing
readers to choose the optimal method for their research.  Moreover, the very
aggressive subtraction (bgsub\_v6) and no subtraction (bgsub\_v0) can be used
as the lower and upper limits of clump fluxes to evaluate the uncertainty
caused by subtracting the diffuse background. The ``fiducial'' method
(bgsub\_v4) is chosen to simplify the paper, focusing on the clump properties.
We will discuss more the effects of background subtraction on clump properties
in Section \ref{prop:bkg}.

\begin{table*}[htbp]
\caption{Diffuse background measurement \label{tb:bkg}}
\hspace*{-0.5cm}
\begin{tabular}{ccccccc}
\hline\hline
Method & Clump Mask Size\footnote{The radius of clump regions masked out when calculating the diffuse background (see green circles in Figure \ref{fig:bkg}). One pixel is 0\farcs06.}  & Background Aperture\footnote{The inner and outer radii of the aperture used to measure the diffuse background (see the magenta circles in Figure \ref{fig:bkg}).} & Median Relative F160W Flux\footnote{Median of clumps' F160W fluxes normalized by those of bgsub\_v4.} & Photo-z $\sigma_{NMAD}$\footnote{NMAD: normalized median absolute deviation. $\sigma_{NMAD} = 1.48 \times (|\Delta z - median(\Delta z)|/(1+z_{spec}))$, where $\Delta z = z_{photo}-z_{spec}$.}$^,$\footnote{Only use clumps with spec-z and $f_{LUV} \geq 0.08$.} & Photo-z outlier\footnote{Outliers are defined as $|\Delta z|/(1+z) > 0.15$.}  & Comment \\
       & (arcsec) & (arcsec) & (normalized) & & & \\
\hline
bgsub\_v6 & 0.18 & 0.18--0.30 & 0.67 & 0.114 & 6.99\% & Very aggressive subtraction \\
bgsub\_v5 & 0.18 & 0.24--0.36 & 0.86 & 0.088 & 6.86\% & \\
bgsub\_v4 & 0.24 & 0.24--0.36 & 1.00 & 0.074 & 6.66\% & Fiducial subtraction \\
bgsub\_v3 & 0.24 & 0.24--0.42 & 1.08 & 0.072 & 6.59\% & \\
bgsub\_v2 & 0.24 & 0.30--0.42 & 1.14 & 0.073 & 6.55\% & \\
bgsub\_v1 & 0.24 & 0.36--0.48 & 1.25 & 0.063 & 6.44\% & Very conservative subtraction \\
bgsub\_v0 & --- &  --  & 2.17 & 0.042 & 5.76\% & No background subtraction\\
\hline
\end{tabular}
\end{table*}

\begin{figure*}[htbp] \center{
\includegraphics[scale=0.5, angle=0]{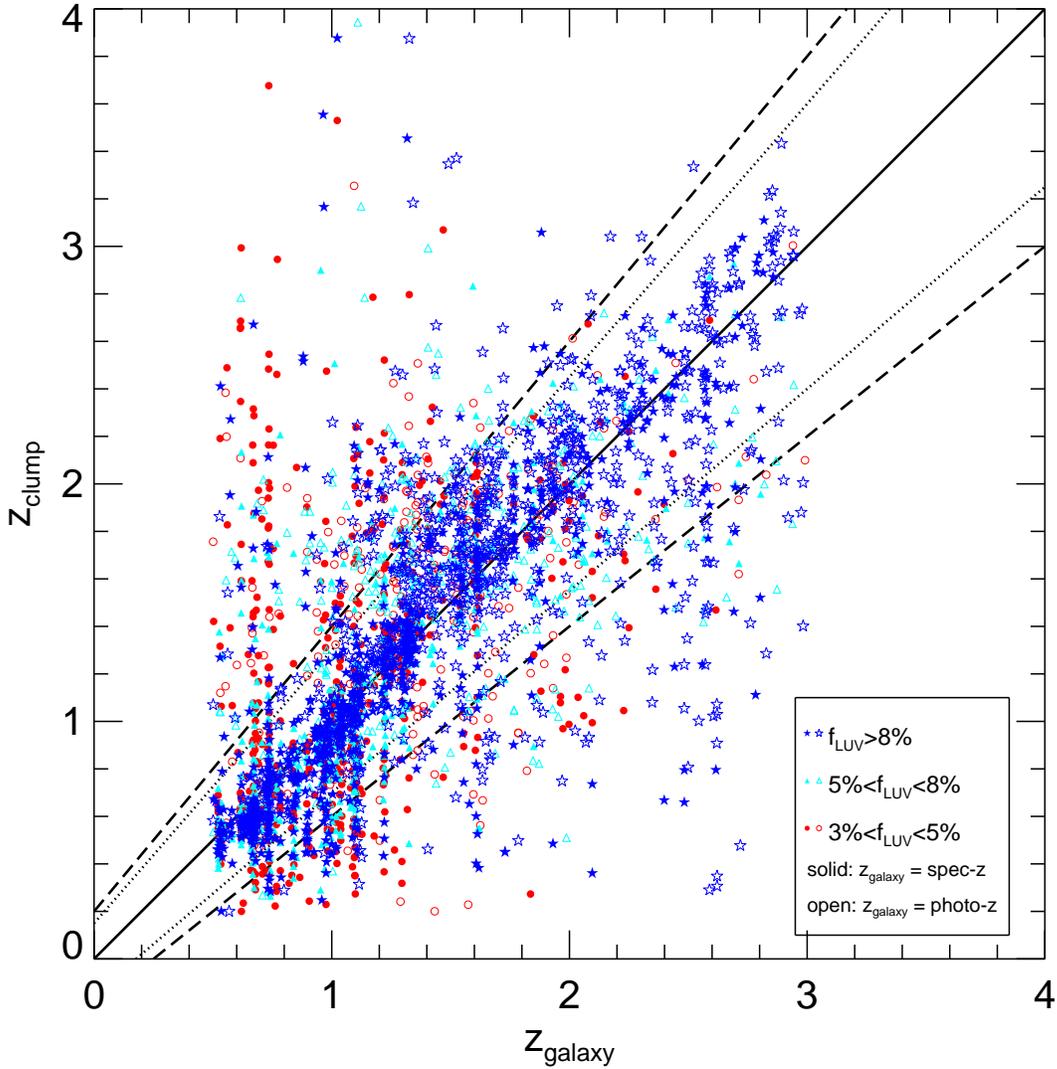}}

\caption[]{Comparison between the photometric redshifts (photo-z) of clumps and
the redshifts of their host galaxies. The photo-z of clumps are measured
through \hst-band PSF-matched photometry (see the text). Our fiducial
background subtraction method bgsub\_v4 (see Table \ref{tb:bkg}) is used. Blue
stars, cyan triangles, and red circles show clumps with UV fractional
luminosity $f_{LUV} \equiv L_{clump}^{UV} / L_{galaxy}^{UV} > 0.08$, $0.05 <
f_{LUV}< 0.08$, and $0.03< f_{LUV} < 0.05$. For each color, the solid symbols
show the clumps whose host galaxies have high-quality spectroscopic redshifts,
while open symbols show the clumps whose host galaxies only have photo-zs from
CANDELS.  The solid line shows the one-to-one correspondence, while the two
dotted lines show $|\Delta z|/(1+z) = 0.15$, which is used to calculate the
outlier fraction. The two dashed lines show $|\Delta z|/(1+z) = 0.2$. Clumps
whose photo-zs are worse than this criterion may have inaccurate multi-band
photometry and hence are excluded from our later analyses.

\label{fig:photoz}}
\vspace{-0.2cm}
\end{figure*}

\subsection{Accuracy of Photometry}
\label{photometry:accuracy}

Background subtraction affects photometric accuracy. An aggressive method
(e.g., bgsub\_v6) subtracts a higher fraction of light from clumps than a
conservative method does. In the former, the remaining clump flux is fainter
and the relative error of the clump flux is therefore larger (or clump S/N is
lower). To test the photometric accuracy, we measure the photo-zs of individual
clumps by using their {\it HST} photometry. The code and details of our photo-z
method are described in \citet{ycguo12vjl}. Models used to measure photo-zs are
extracted from the library of PEGASE 2.0 \citep{pegase}. For integrated
galaxies, our photo-z code achieves a similar accuracy as those used in
official CANDELS photo-z catalogs \citep{dahlen13}. To test the photo-z
accuracy of clumps, we only use clumps with spectroscopic redshifts and $f_{UV}
> 8\%$.

Table \ref{tb:bkg} shows that when the background subtraction is too aggressive
(e.g., bgsub\_v6), the scatter (the normalized median absolute deviation
$\sigma_{NMAD}$) and outlier (defined as sources with $|\Delta z|/(1+z)>0.15$)
fraction of photo-z measurement are large. In contrast, conservative (or no)
subtraction (bgsub\_v1 or bgsbu\_v0) yields much improved photo-z statistics. 

The photo-z result of the fiducial subtraction method is shown in Figure
\ref{fig:photoz}. The $\sigma_{NMAD}$ (normalized median absolute deviation)
and outlier fraction of bgsub\_v4 are 0.074 and 6.66\%, both of which are about
a factor of 2.5 larger than the values of CANDELS official photo-z catalogs of
\citet{dahlen13}. The worse photo-z accuracy of clumps is expected, because (1)
clumps are much fainter than integrated galaxies (e.g., see Figure
\ref{fig:massmag}); (2) clumps only have photometry of a few {\it HST} bands,
while integrated galaxies usually have more than 15 bands; (3) clump photometry
only samples the relatively featureless regime of SEDs of star-forming
population models i.e., rest-frame UV--optical without a strong Balmer/D4000
break; and (4) background subtraction itself induces a source of uncertainty.
Considering all these factors, the photo-z accuracy is acceptable. In this
paper, clump photo-zs are only used for the purpose of testing their
photometry. They are not used for deriving clump properties. We use the
redshifts of the host galaxies to derive clump properties.

Some clumps have catastrophic photo-z measurements with $|\Delta z|/(1+z)>0.20$
(dashed lines in Figure \ref{fig:photoz}), which may indicate problematic
photometry. We exclude such clumps in our later analyses, although still keep
them in the published catalog. We set a flag {\it badczflag = 1} in the catalog
to label these clumps. In total, 720 (out of total 3193) clumps are thereby
excluded. Among the 720 clumps, 264, 187, and 269 clumps have UV
fractional luminosity $f_{LUV} \geq 0.08$, $0.05 \leq f_{LUV} < 0.08$, and
$0.03 \leq f_{LUV} <0.05$, respectively.

\section{Measuring Clump Properties}
\label{measureprop}

\subsection{Spectral Energy Distribution Fitting}
\label{measureprop:sed}

We derive the physical properties (\mstar, SFR, age, and dust reddening) of
clumps by fitting their {\it HST} SEDs to stellar population synthesis models
retrieved from the library of \citet[][]{bc03} with a Chabrier IMF
\citep{chabrier03}. The details of our SED-fitting code are described in
\citet{ycguo12vjl}. Briefly, we use a set of $\tau$-models in which star
formation history declines exponentially with time. The set of models consists
of grid points in a parameter space spanned by redshift, dust extinction
E(B-V), star formation history (SFH) characterized by $\tau$ and age, and
metallicity. The available values of each parameter are shown in Table
\ref{tb:sedparam}. We apply the Calzetti extinction law \citep{calzetti97,calzetti00} and
the recipe of \citet{madau95} to the models to account for dust extinction and
the opacity of the IGM in the universe. We use the minimal $\chi^2$ value to
decide the best-fit model.  During the SED-fitting, the redshift of a clump is
fixed to that of its host galaxy (spec-z if available, photo-z otherwise). For
each clump, we Monte Carlo sample its photometry in each band 100 times from a
Gaussian distribution whose mean is equal to the observed flux and whose
standard deviation is equal to the flux uncertainty. We then fit the 100
re-sampled SEDs. For each stellar population parameter, the average of the 100
best-fit values is used as the best value, and the 16th and 84th percentiles of
the 100 best-fit values are used as the 1$\sigma$ confidence level.

\begin{table}[htbp]
\caption{Parameter Space of SED-Fitting \label{tb:sedparam}}
\begin{tabular}{ccc}
\hline\hline
Parameter & &  Range \\
\hline
Redshift & & 0.0 to 7.0 with a bin size of 0.01 \\
E(B-V)\footnote{E(B-V) runs up to 0.3 for models with $t/\tau>=4.0$.} & & 0.0 to 1.0, $\Delta E(B-V)=0.05$ \\
Metallicity & & solar \\
Age (Gyr)\footnote{Age is defined as the period from the onset of star formation (i.e., the beginning of the $\tau$-model or constant constant SFH model) to the time the object is observed.}  & & (1, 2, 3, 5, 8) $\times$ $10^{-2}, 10^{-1}, 10^{0}, 10^{1}$, up to 13 \\
$\tau$ (Gyr) & & (1, 2, 3, 5, 8) $\times$ $10^{-1}, 10^{0}, 10^{1}$, and $\infty$\footnote{$\tau=\infty$ means a constant SFH.} \\
\hline
\end{tabular}
\end{table}

\begin{figure*}[htbp] \center{
\hspace{-0.2cm}
\includegraphics[scale=0.4, angle=0]{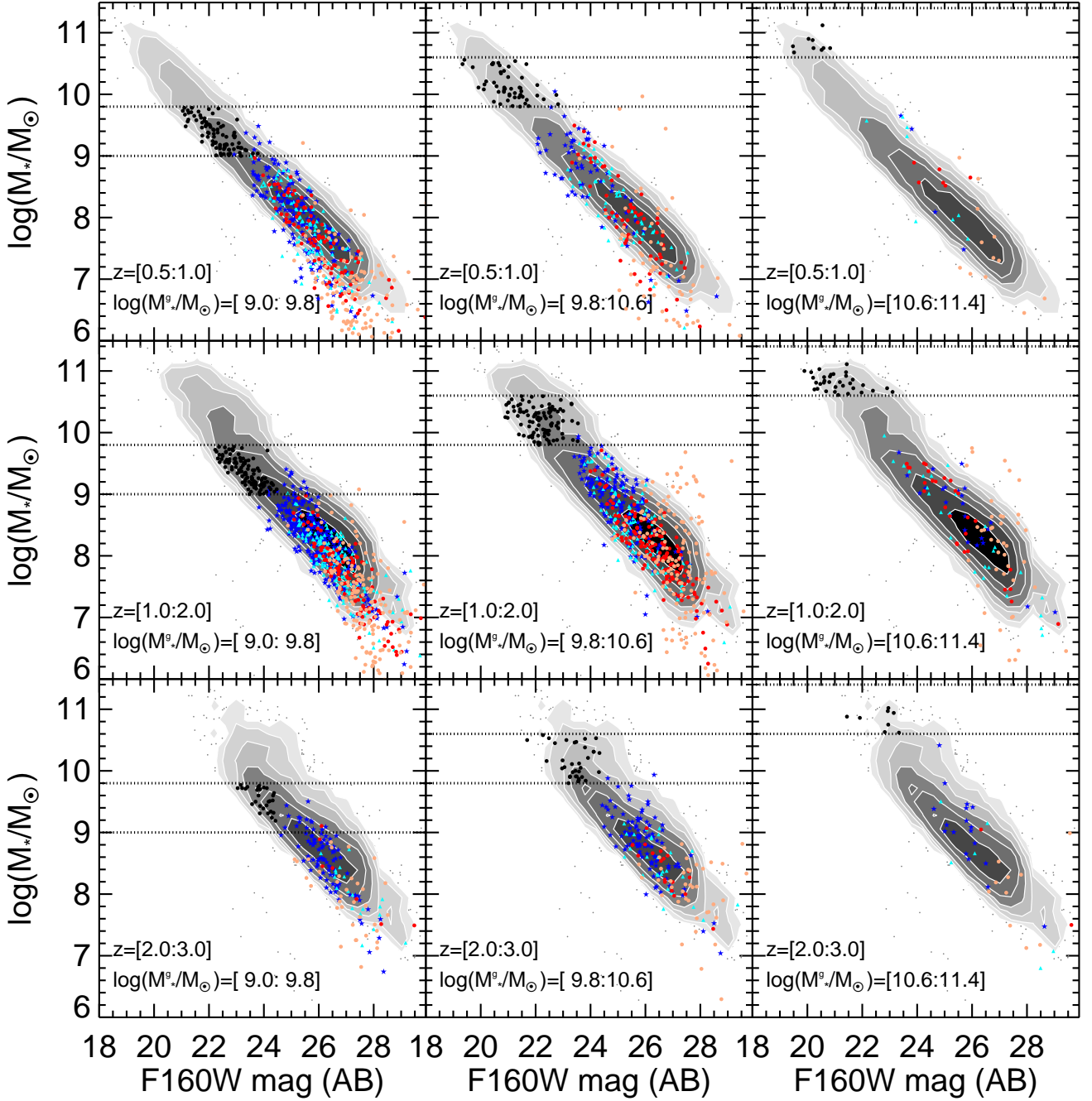}}

\caption[]{\mstar--magnitude (F160W) diagram of clumps. Each panel shows clumps
detected from galaxies within a given \mstar\ and redshift bin as the labels
indicate. Similar to Figure \ref{fig:photoz}, in each panel, blue, cyan, and
red clumps have $f_{LUV} \equiv L_{clump}^{UV} / L_{galaxy}^{UV} > 0.08$, $0.05
< f_{LUV}< 0.08$, and $0.03< f_{LUV} < 0.05$, respectively. The orange clumps
are those excluded because of their large photo-z errors $|\Delta z|/(1+z) >
0.2$. The gray contours and gray points show the distribution of all
CANDELS/GOODS-S galaxies (not just our sample galaxies) within the same
redshift range of each panel but simultaneously for all mass ranges.
In each panel, the two horizontal dotted lines show the \mstar\ range of the
host galaxies. The black dots in each panel show the host galaxies of detected
clumps.

\label{fig:massmag}}
\vspace{-0.2cm}
\end{figure*}

SED-fitting results significantly rely on the assumed SFHs. Unfortunately, the
SFHs of clumps are little known. Here, we choose the $\tau$-model and constant
SFH model because they are commonly used in the SED-fitting of distant
galaxies. Recently, other models, e.g. inverse $\tau$-model and delayed
$\tau$-model are also frequently used in the literature. An unrealistic SFH
model would result in systematic errors of the derived parameters, especially
for age. Lee et al. (2017, in preparation) test the effects of different models
on the derived parameters for integrated CANDELS galaxies. They generate mock
SEDs with known intrinsic parameters and fit the SEDs with different SFH
models.  They find that (1) \mstar\ is the most robust parameter and nearly
unaffected by the adapted models; (2) constant SFH recovers age (defined as the
period from the onset of star formation to the time the object is observed)
with little systematic offset, while $\tau$-model underestimates age by
$\sim$0.3 dex (for a system of $\lesssim$ 1 Gyr old); and (3) for systems with
SFR$>$0.01\msun/yr, constant SFH overestimates SFR by 0.25 dex, while
$\tau$-model recovers SFR with little systematic offset. We expect similar
systematic errors due to SFH assumptions in our clump SED-fitting. However,
since our data cover a much shorter wavelength range and have larger
photometric uncertainties, the random errors of our clump SED-fitting are
larger than integrated galaxy SED-fitting. We do not find any systematic trend
between clumps age and the SED-fitting preferred SFHs.

\subsection{Tests of Clump Properties: I. Mass-to-light Ratio}
\label{measureprop:ml}

We test the accuracy of clump properties in three ways. The first is the
mass-to-light ratio ($M/L$). Among all {\it HST} bands, F160W is the reddest
and therefore serves as the best \mstar\ indicator in our method. We expect
that $M/L$ of clumps is similar to or slightly smaller than that of integrated
galaxies at a given \mstar, because clumps are believed to be younger. In
Figure \ref{fig:massmag}, we plot the relation between \mstar\ and F160W
magnitude for clumps and integrated CANDELS/GOODS-S galaxies. In each panel, we
plot clumps that are detected from galaxies within a given \mstar\ and redshift
range (as shown by the label and dashed vertical lines in each panel), but plot
all CANDELS/GOODS-S galaxies (not just our sample galaxies), regardless
of their \mstar, within the redshift range.  

Clumps follow a similar relation with integrated galaxies. This result is
especially true at $0.5\leq z<1.0$, when F160W is sampling $\sim$9000\AA, very
close to the peak of stellar emission. In this redshift range, the variation of
$M/L$ caused by different stellar populations is the smallest. At higher
redshifts, the $M/L$ of clumps is slightly smaller than that of integrated
galaxies, demonstrated by the fact that clumps are lying toward the lower
boundary of the galaxy \mstar--magnitude relation. The difference could be
caused by a bandpass coverage effect. At $z\geq1$, {\it HST} filters are
shifted to cover bluer side of the SED than at $z<1$, probing more the young
star component and less the old star component. As a consequence, our clump
mass measurement is likely weighting more the UV luminosity and artificially
biasing the $M/L$ to lower values. The difference, however, is small, because
galaxies with \mstar\ similar to that of clumps at $z>1$ are also very actively
forming stars. Overall, this test shows that there are no obvious, significant
systematics in the \mstar\ measurement of clumps. Also, the large scatter of
clumps with bad clump photo-z (i.e., those with {\it badczflag = 1}, orange
circles in the figure) supports our decision of not including them in later
analyses, because their problematic photometry causes large uncertainties in
\mstar. 

\begin{figure*}[htbp] \center{
\hspace{-0.2cm}
\includegraphics[scale=0.4, angle=0]{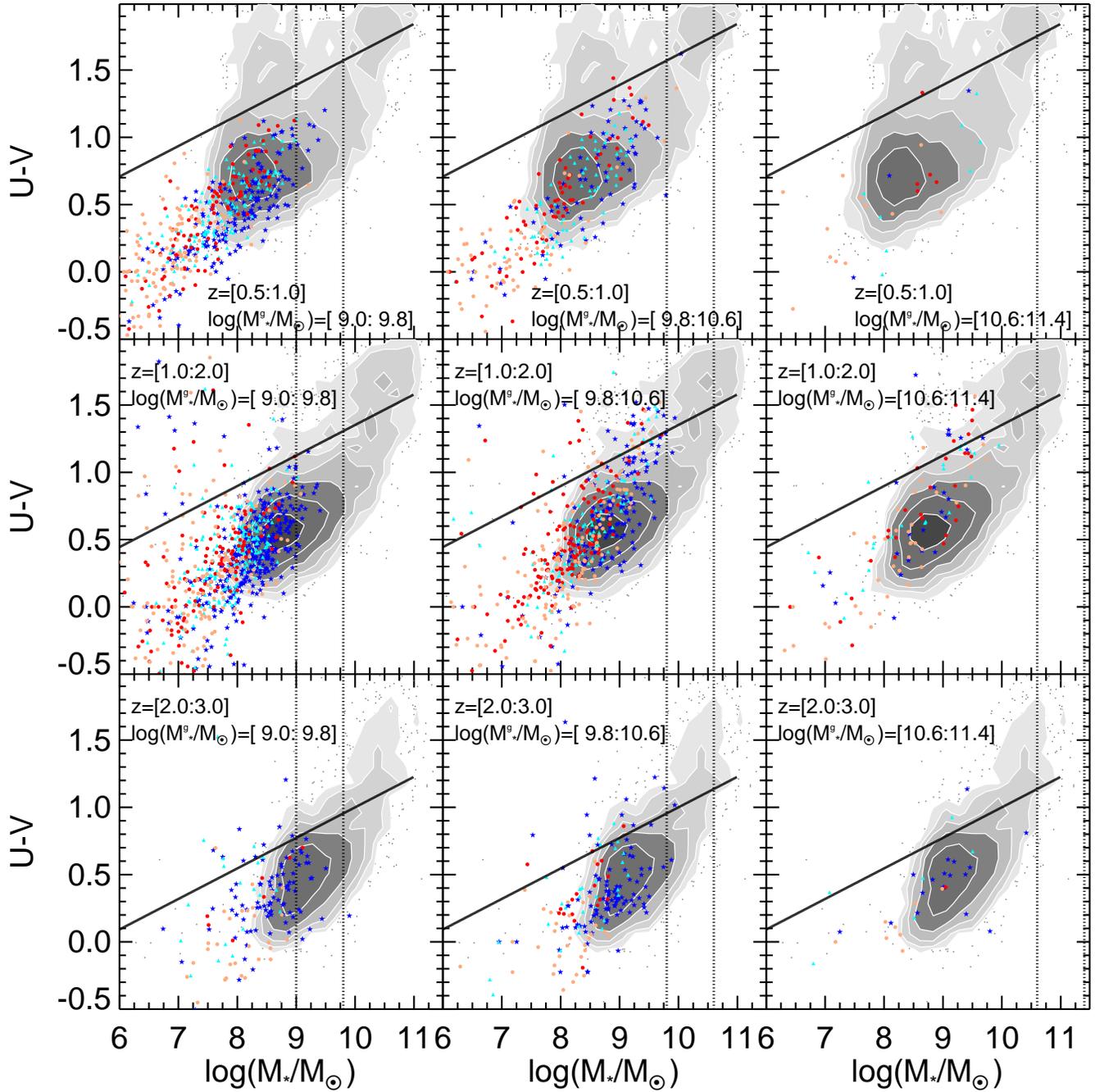}}

\caption[]{Similar to Figure \ref{fig:massmag}, but showing the color--\mstar\
diagram of clumps (color symbols) and CANDELS/GOODS-S galaxies (gray contours
and points). 
The solid line in each panel is the separation between red-sequence and blue
cloud derived in \citet{borch06}. In each panel, the two vertical dotted lines
show the \mstar\ range of the host galaxies where clumps are identified (see
also the \mstar\ label).

\label{fig:colormass}}
\vspace{-0.2cm}
\end{figure*}

\subsection{Tests of Clump Properties: II. Color--mass Diagram}
\label{measureprop:cm}

The second test is the color--\mstar\ diagram. In Figure \ref{fig:colormass},
we plot both clumps and integrated galaxies in the diagram of rest-frame U-V
vs.  \mstar. Clumps have similarly blue colors as those galaxies whose \mstar\
is comparable to the clump \mstar, but clumps are bluer than their host
galaxies. This result is expected since clumps are selected as UV-bright
regions from their galaxies. At a given redshift and galaxy \mstar, clump
colors show a relation with their \mstar: massive clumps are redder than
lower-mass clumps. Also, for a given clump mass, UV-bright clumps (blue dots)
are bluer than UV-faint clumps (red dots), as expected. We also notice that
some clumps are as red as red-sequence galaxies, i.e., above the separation
line of the blue cloud and the red sequence of integrated galaxies from
\citet{borch06}: $(U-V) = 0.227 \times log(M_{*}/M_\odot) - 1.16 - 0.352 \times
z + (0.79-0.02)$, where the last term is to convert the Vega magnitude in
\citet{borch06} to AB. The very red colors may indicate problematic photometry,
because our clump identification is designed to select star-forming (and
implicitly low dust extinction) clumps. We therefore exclude all clumps above
the separation line (solid black lines in Figure \ref{fig:colormass}) in our
later analyses. In the catalog, we set a flag {\it veryredflag=1} to label
them. The fraction of the ``veryredflag=1'' clumps increases with redshifts:
$\sim$1\% at $0.5 \leq z < 1.0$, $\sim$7\% at $1.0 \leq z < 2.0$, and $\sim$8\%
at $2.0 \leq z < 3.0$.  This result is likely caused by the filter coverage: at
higher redshift, {\it HST} filters are shifted to cover bluer wavelengths, 
resulting in large uncertainties in the measure of the rest-frame V-band
luminosity. In total, 142 clumps are thereby excluded (after {\it badczflag=1}
applied), reducing the clean sample to 2331 clumps.

\begin{figure*}[htbp] 
\hspace{-0.2cm}
\includegraphics[scale=0.3, angle=0]{fig6a.ps}
\includegraphics[scale=0.3, angle=0]{fig6b.ps}

\caption[]{Comparisons of SFRs (left) and dust extinction E(B-V)s (right)
measured through SED-fitting and rest-frame UV continuum. The blue, cyan, red
data points are clumps with $f_{LUV} \equiv L_{clump}^{UV} / L_{galaxy}^{UV} >
0.08$, $0.05 < f_{LUV}< 0.08$, and $0.03< f_{LUV} < 0.05$, respectively. Black
circles with errorbars show the median and 16th and 84th percentiles of the
data.
Clumps at $z\lesssim1.5$ do not have enough {\it HST} filters to sample their
rest-frame UV continuum, and therefore are not shown in the figure.

\label{fig:sfrtest}}
\vspace{-0.2cm}
\end{figure*}
 
\subsection{Tests of Clump Properties: III. SFR and E(B-V)}
\label{measureprop:sfr}

The third test is clump SFR and dust extinction E(B-V). In addition to the
SED-fitting, we also estimate the SFRs and E(B-V)s of clumps by using the slope
and luminosity of their rest-frame UV continuum. Compared to SED-fitting, this
method is less model-dependent and requires no prior information of the SFH of
galaxies. We first calculate the UV slope through a linear fit of 
$log(f) \propto \beta \lambda$, where $f$ is flux, $\lambda$ wavelength, 
and $\beta$ the UV slope.
We then use the Calzetti extinction law \citep{calzetti94, calzetti00} to convert the
rest-frame UV-slope of a clump into its dust reddening E(B-V), and calculate
the unobscured SFR from its dust-corrected rest-frame UV continuum by using the
formula in \citet{kennicutt98}, which applies to systems with constant star
formation over timescales of $\sim$100 Myrs. The rest-frame UV continuum used
here covers the wavelength range of 1400\AA\ to 2800\AA. We require the clumps
to have at least two {\it HST} bands to sample this range. Under this
requirement, the bluest {\it HST} band in our dataset (F435W) only enables the
measurement of UV SFR for clumps at $z\gtrsim1.5$. Therefore, although UV SFR
is less model-dependent than SED-fitting-derived SFR, we only use the former to
test the latter. For the whole clump sample, we still use SED-fitting-derived
SFRs as our measurement.

Figure \ref{fig:sfrtest} shows very good agreement between SED-fitting and UV
continuum derived SFRs and E(B-V)s. The average difference between SED SFR and
UV SFR (${\rm \Delta SFR = log(SFR_{SED})-log(SFR_{UV})}$) is about 0.07 dex
and the 1$\sigma$ scatter of ${\rm \Delta SFR}$ is about 0.4 dex -- slightly
larger than the typical SFR uncertainty for integrated galaxies in the
literature. For E(B-V), the average difference is about 0.02 and the 1$\sigma$
scatter is about 0.1. The agreement in the higher redshift range $2.0<z<3.0$ is
better than that in the lower range $1.5<z<2.0$, because in the former, three
or more {\it HST} bands are sampling the rest-frame UV continuum, enabling a
more accurate measurement, while in the latter, only two {\it HST} bands are
available. 

Overall, all the above tests find no obvious and significant systematics in our
measurements of clump properties. Some major stellar population parameters ---
\mstar, SFR, and E(B-V) --- are measured to a reasonable accuracy level. All
these tests also provide criteria to exclude problematic clumps from the
sample. As a summary, we exclude clumps that have (1) bad photo-z ({\it
badczflag=1}) or (2) very red U-V color ({\it veryredflag=1}).

\begin{figure*}[htbp] \center{
\includegraphics[scale=0.4, angle=0]{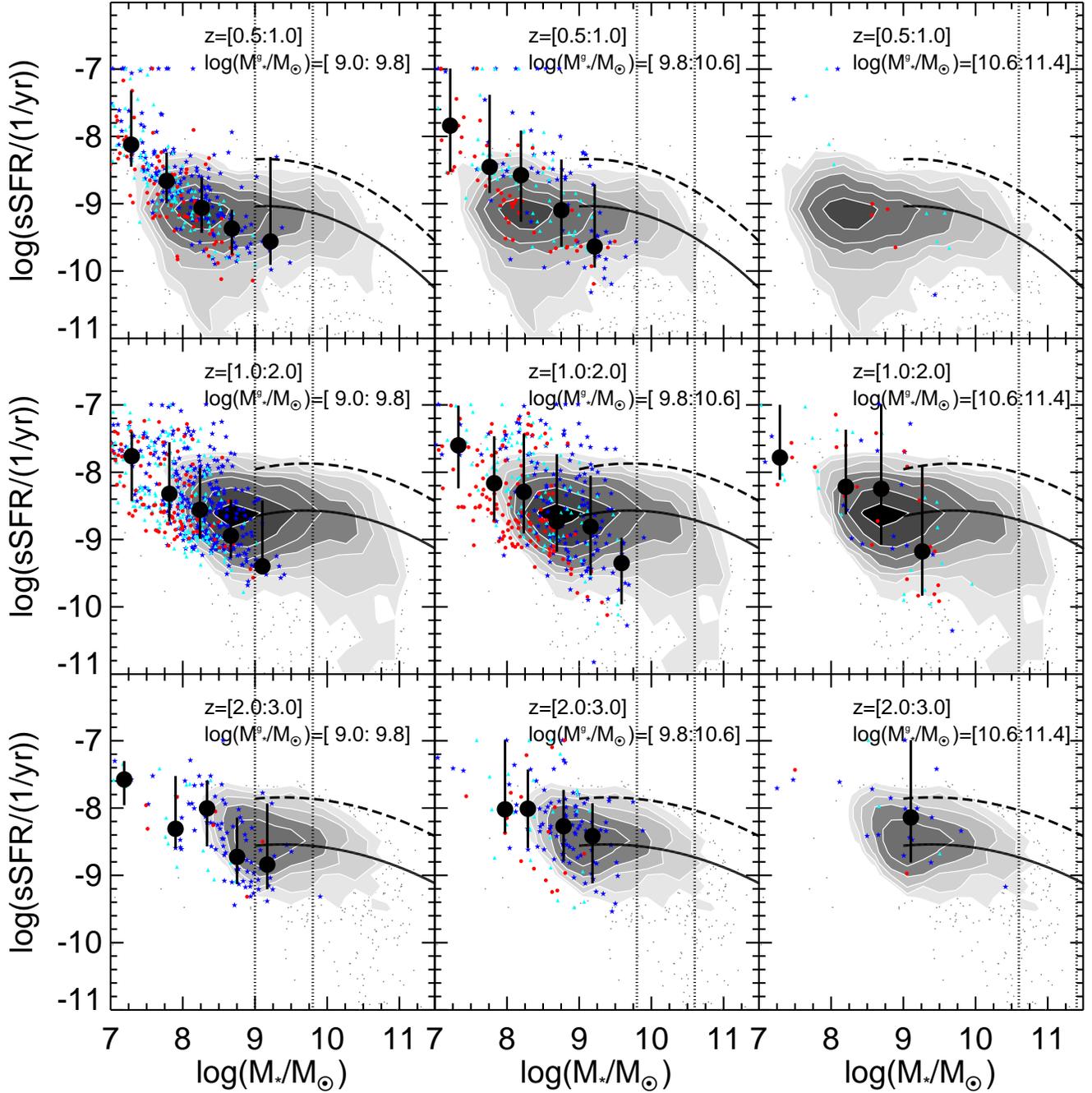}}

\caption[]{Similar to Figure \ref{fig:colormass}, but showing the sSFR--\mstar\
relation of clumps and integrated CANDELS/GOODS-S galaxies with F160W$<$26 AB.
Color symbols, gray contours and points, and vertical dotted lines have the
same meaning as in Figure \ref{fig:massmag} and \ref{fig:colormass}. Black
circles with errorbars show the median and 16th and 84th percentiles of clump
sSFRs in individual \mstar\ bins. The solid black curves show the polynomial
fit of the star formation sequence of \citet{whitaker14}, and the dashed black
lines are the solid lines scaled up by a factor of five. 

\label{fig:ssfrms}}
\vspace{-0.2cm}
\end{figure*}

\subsection{Tests of Clump Properties: IV. Comparison with Integrated Values}
\label{measureprop:global}

We also carry out two additional sanity checks to use the integrated values as
constraints on clump properties. For each galaxy, (1) the total \mstar\ in
clumps should be lower than the integrated \mstar\ of the host galaxy and (2)
the total SFR in clumps should be smaller than the integrated SFR of the host
galaxy. These tests provide additional information on the robustness of
results. We only use the clean sample of 2331 clumps (see Section
\ref{measureprop:cm}) for this test.

In our fiducial background subtraction method (bgsub\_v4), 13 galaxies (1\% of
the total 1270 galaxies) fail the total clump mass test, while 32 galaxies (3\%
of the total galaxies) fail the total SFR test. The reasons of failure,
however, are different between the mass and SFR tests. Among the 13
galaxies failed the mass test, only two of them have a single clump that is
more massive than the integrated galaxy (such clumps are flagged with {\it
badmassflag}=1 in the catalog). In the SFR test, however, 21 of 32 failed
galaxies each contains a single clump with SFR larger than the integrated SFR
of the galaxies. If we exclude these 21 clumps with over-estimated SFR, only 11
galaxies (1\%) fail the SFR test. The failure fraction also depends on
the background subtraction, with aggressive subtraction resulting in less
failure galaxies or clumps.

\section{Observed Clump Properties}
\label{observedprop}

In this section, we present some observed clump properties that may be of
interest to readers. We only present these results and discuss their
systematics and uncertainties. The theoretical interpretations and implications
of the observed clump properties are beyond the scope of this paper and are
thereby left for future work.  
In most figures in this section, we divide our sample into different redshift
and galaxy \mstar\ bins (same as in Figures \ref{fig:massmag} and
\ref{fig:colormass}). The symbols and colors in these figures are also the same
as in Figures \ref{fig:massmag} and \ref{fig:colormass} unless otherwise
stated. All these properties are measured with our fiducial diffuse background
subtraction bgsub\_v4 (see Table \ref{tb:bkg}).

\subsection{Specific Star Formation Rate vs. \mstar}
\label{prop:ssfr}

Figure \ref{fig:ssfrms} shows the relation between sSFR and \mstar\ for clumps
(color symbols) and integrated CANDELS/GOODS-S galaxies with F160W$<$26 AB
(gray contours and points). 
In most panels, the sSFR of clumps increases with the decrease of clump \mstar.
In contrast, galaxies below $10^{10}$\msun\ have almost constant sSFR evident
by the almost horizontal gray contours. 
On average, when clumps' \mstar\ is significantly lower than their host
galaxies' \mstar, the sSFR of clumps is about 5 times higher than the typical
sSFR in galaxies with \mstar\ similar to their host galaxies (i.e., compare
massive clumps with contours within the two vertical dotted lines). This result
is consistent with other studies \citep[e.g.,][]{ycguo12clump,wuyts12}. 
The clear trend is for clump's sSFR to increase with the decrease of \mstar,
with the most massive clumps having sSFR similar to that of their host
galaxies. Very low-mass clumps' sSFR is up to 30 times higher than that of the
host galaxies. 

\begin{figure*}[htbp] \center{
\includegraphics[scale=0.4, angle=0]{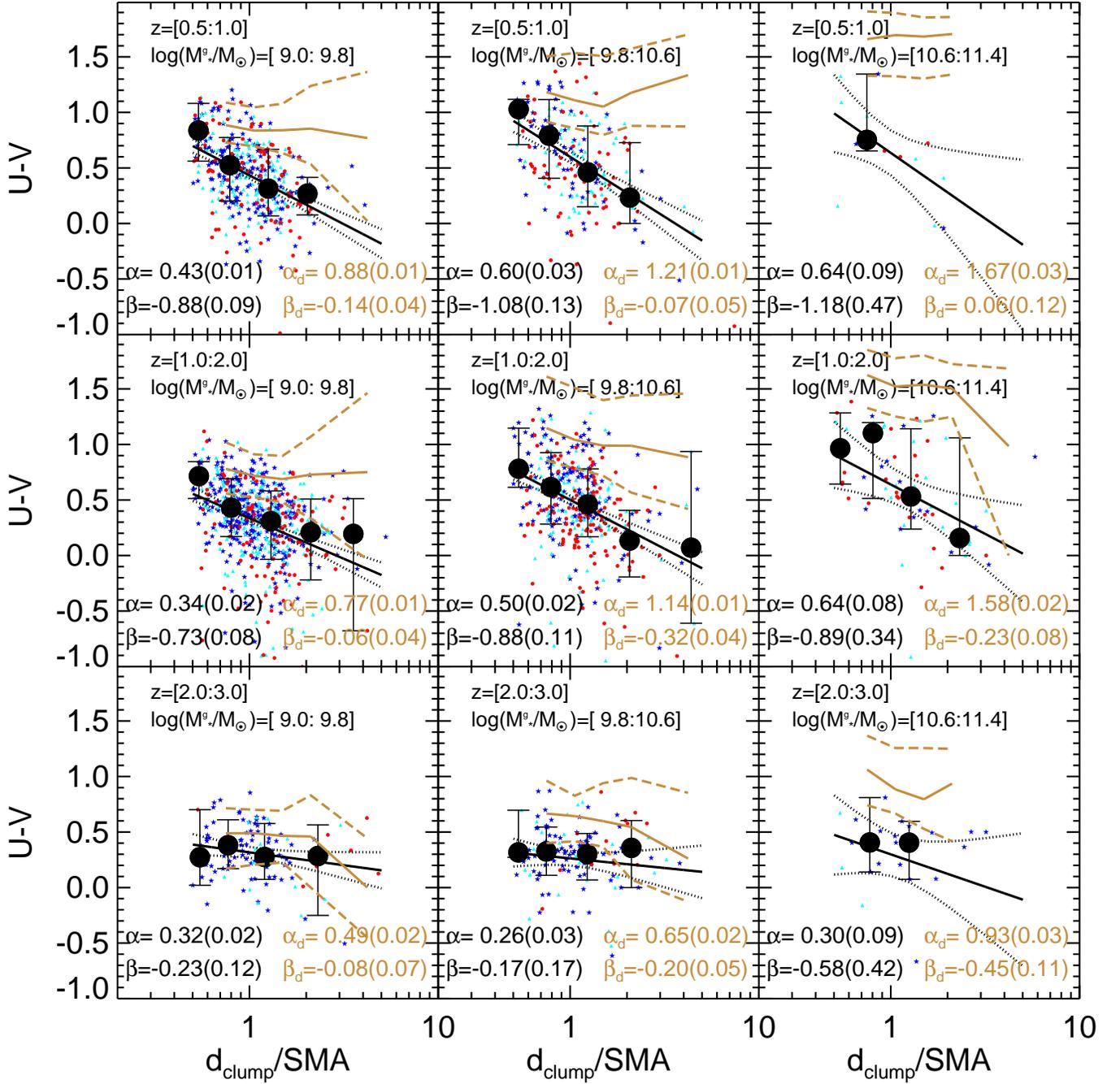}}

\caption[]{Variation of the rest-frame U-V color of clumps as a function of
clumps' galactocentric distance normalized by SMA from host galaxy centers.
Blue, cyan, and red points and black circles with errorbars are the same
as in Figure \ref{fig:massmag}, \ref{fig:colormass}, and \ref{fig:ssfrms}. The
galactocentric distance of clumps ($d_{clump}$) is scaled by the semi-major
axis ($SMA$) of their galaxies.  In each panel, the black solid line and dotted
curves show the best linear fit ($(U-V) = \alpha + \beta log(d_{clump}/SMA)$)
and its confidence level to the color points. The intersection ($\alpha$) and
slope ($\beta$) of each linear fit are shown in the lower left corner of each
panel. The values within the parentheses are errors. Solid and dashed light
brown curves in each panel show the median and deviation of the radial gradient
of the U-V color of intra-clump regions. The $\alpha$ and $\beta$ of the best
linear fit to the diffuse component gradient is shown in the lower right
corner.

\label{fig:clgrad_color}}
\vspace{-0.2cm}
\end{figure*}

\subsection{Radial Variation of Clump Color (Color Gradient)} 
\label{prop:clgrad_color}

Figure \ref{fig:clgrad_color} shows the variation of the clump U-V color as a
function of galactocentric distance normalized by the SMA of their host
galaxies (color gradient). 
The clump age gradient and sSFR gradient are usually used to test theoretical
models of clump formation and evolution. These quantities, however, can only be
derived through colors (SEDs) in our dataset. Therefore, we first present the
color gradient, because it is directly observed and the most robust result
among all gradients discussed in this paper.

We find a color gradient in almost all panels at $z<2$: clumps at small
galactocentric distance (normalized by the SMA of their galaxies) are redder,
while those at large distance are bluer. This result is similar to many other
studies in the literature: e.g.,
\citet{fs11b,ycguo12clump,tadaki14,shibuya16,soto17}. At $z\geq 2$, clumps only
show almost no color gradients.

Our large dataset covering a wide redshift and \mstar\ ranges enable us to
study the dependence of the color gradient on redshift and galaxy \mstar\ for
the first time. At the same galaxy \mstar, the color gradient becomes steeper
toward lower redshifts. Also, at the same redshift, the gradient becomes {\it
slightly} steeper toward more massive galaxies.

\begin{figure*}[htbp] \center{
\includegraphics[scale=0.4, angle=0]{fig9.ps}}

\caption[]{Similar to Figure \ref{fig:clgrad_color}, but for the radial
gradient of the age of clumps. 

\label{fig:clgrad_age}}
\vspace{-0.2cm}
\end{figure*}

We also calculate the color gradient of intra-clump regions (or diffuse
background). To this purpose, we measure the multi-band photometry of the host
galaxies in circular annuli after masking the clump regions. We then use the
same method in Section \ref{measureprop} to derive the physical properties of
intra-clump regions. 
Figure \ref{fig:clgrad_color} shows that, overall, (1) intra-clump regions
(brown lines in the figure) are redder than clumps and (2) the color gradient
of intra-clump regions is flatter than that of clumps. In the lowest-\mstar\
bin ($10^9-10^{9.8}$\msun), the intra-clump regions' color gradient is almost
flat in all redshift bins. We also find marginal evidence (through a linear
fit) that the slope of intra-clump regions' color gradient becomes steeper with
galaxy \mstar\ at $z\geq 1$. This result is consistent with recent studies of
the color gradient of integrated light in galaxies, e.g.,
\citet{liufs16,wangwc17,tacchella17}.

Figure \ref{fig:clgrad_color} implies that clumps' color gradient at $z<2$
would be changed, if the colors were measured by using the fluxes without
background subtraction. The color gradients of diffuse background (brown solid
lines) are significantly different from that of clumps (large black circles):
the former is almost flat, while the latter decreases with the normalized
galactocentric distance. Therefore, if the measurements without background
subtraction are used, clumps would be redder than with background subtraction,
with a big impact especially on the age--extinction determination.  Moreover,
without background subtraction, the clump color gradient would be flatter (see
the top left panel of Figure \ref{fig:aper} and more discussion in Section
\ref{prop:bkg}). This result indicates that color gradients of clumps can only
be made shallower -- not the opposite -- by the contribution of the diffuse
background. It is hard to justify the observed color gradients in clumps as
being introduced only by background contamination, which in the worst case
would ``dilute'' those gradients.  Therefore, the existence of a negative color
gradient
can be considered quite robust with respect to background contamination (or
background subtraction method).

On the other hand, since the individual clump contributes only a few percents
to the total UV light of their galaxies and even less to \mstar\ of the
galaxies, the global color gradient of galaxies is actually dominated by the
diffuse component, which is supported by the broad consistency between our
measurement of the color gradient of diffuse component and other measurements
of global color gradient in the literature.

The color gradient in Figure \ref{fig:clgrad_color} is measured when the
(projected) galactocentric distance is normalized by the semi-major axis of the
galaxies. We also use the physical projected galactocentric distance (in unit
of kpc) to measure the color gradient. Qualitatively, all the above results are
not changed. The slopes of the clump color gradient using the physical distance
are actually steeper than those using the normalized distance, except in the
lowest-mass bin at the highest redshift. We keep using the normalized
galactocentric distance for other gradients below.

Figure \ref{fig:clgrad_color} also provides tests on the two selection effects
discussed in Section \ref{sample:bias}: redshift dependent and galaxy \mstar\
dependent biases, both introduced by our relative definition of clumps with a
fixed \fluv\ (i.e., at higher redshifts or higher galaxy \mstar, only high
\fluv\ clumps are detected).
For clumps in galaxies with \mstar$<$10.8 at $z<2.0$ (where we have enough
clumps), we re-calculate the clump U-V color gradient by dividing clumps into
three sub-samples: 0.03$\leq$\fluv$<$0.05, 0.05$\leq$\fluv$<$0.08, and
\fluv$\geq$0.08. The three sub-samples in each ($z$, \mstar) bin show almost
the same gradient, indicating that our results have almost no dependence on the
adopted \fluv\ thresholds.
The comparisons between different galaxy \mstar\ bins at a given redshift is
also robust with respect to the \fluv\ threshold. For example, the result of
the clump color gradient slopes increasing with galaxy \mstar\ is still true
even when we match the intrinsic luminosity of clumps in different galaxy
\mstar\ bins (e.g., by comparing low \fluv\ clumps from galaxies with
\mstar$>$9.8 with high \fluv\ clumps from galaxies with \mstar$\leq$9.8).
Overall, we conclude that our results of clump color gradient are not
significantly affected by the selection effects.

\subsection{Age Gradient}
\label{prop:clgrad_age}

An important test of different clump evolution models is the age gradient. The
inward migration scenario
\citep[e.g.,][]{bournaud07,elmegreen08,ceverino10,mandelker14,bournaud14a}
predicts a negative age gradient: inner (small galactocentric distance) clumps
are older, while outer (large galactocentric distance) clumps are younger. In
these models, clumps spend a few hundred Myr migrating from galaxy outskirts to
galactic centers. Therefore, the age difference between inner and outer clumps
should also be on the order of a few hundred Myr. Such a negative age gradient
is found by some observations \citep[e.g.,][]{fs11b,ycguo12clump,soto17},
consistent with the inward migration scenario. A few simulations, e.g., FIRE
\citep{oklopcic17} and NIHAO \citep{buck16}, however, argue that the age
gradient may be a result of clumps being contaminated by old disk stars that
happen to be in clump locations.  Although these simulations are able to
reproduce the trend of the observed clump age gradient, clump migration is not
found in them. The clump ages in FIRE are significantly shorter -- less than 50
Myr. In this paper, as discussed in Section \ref{photometry}, we try different
diffuse background subtraction configurations to {\it statistically} minimize
the contamination of ``disk'' stars. 

Figure \ref{fig:clgrad_age} shows clump age as a function of clump
galactocentric distance scaled by SMA. In our SED-fitting, age is defined as
the period from the onset of star formation (i.e., the beginning of the
exponentially declining $\tau$-model or constant SFH model) to the time the
object is observed. The existence of a clump age gradient depends on the
redshift and \mstar\ of their host galaxies. We fit the relation $log(age) =
\alpha + \beta \times log(d_{clump}/SMA)$ to our clump data, and use $\beta$
and its uncertainty (values are shown by black text in the figure) to determine
if an age gradient is significant. For galaxies with \mstar$<10^{10.6}$\msun\
and $z<2.0$, $\beta$ is smaller than zero by more than 3$\sigma$. We therefore
conclude an existence of clump age gradients for these galaxies. For galaxies
at $z\geq2.0$, $\beta$ is consistent with zero within $\sim$1$\sigma$,
indicating a flat age distribution with galactocentric distance, namely no
gradient. For very massive (\mstar$\geq10^{10.6}$\msun) galaxies at $z<2.0$,
$\beta$ deviates from zero by about 2$\sigma$, showing a marginal age
gradient. Given the small number statistics of very massive galaxies, no firm
conclusion can be drawn from our dataset for them. Future studies of larger
samples are needed.

Using the best-fit relation, we can calculate the age difference between
inner and outer clumps. We use $d_{clump}/SMA=0.5$ as the typical location
of inner clumps. At a distance smaller than this, we cannot separate clumps
from galactic bulges due to the resolution of {\it HST} images (see Paper I for
related discussions). For outskirts, we use $d_{clump}/SMA=2.0$ as the typical location
of the outer clumps. This choice is motivated by the fact that, for star-forming
galaxies with \sersic\ index $n=1$, the disk size is about 2$\times$SMA. For
galaxies with \mstar$<10^{10.6}$\msun\ and $z<1.0$, the age difference between
$d_{clump}/SMA=0.5$ and $2.0$ is about 700 Myr, while for galaxies with
\mstar$<10^{10.6}$\msun\ and $1.0\leq z < 2.0$, the age difference is about
250--300 Myr. 

We also compare clump age gradients with the age gradients of diffuse
background (brown lines and brown values in the figure). The diffuse background
(or ``disk'') properties are measured from the annulus photometry as described
in Section \ref{prop:clgrad_color}. We use the same SED-fitting procedure to
derive the properties, e.g., age, \mstar, SFR, E(B-V), etc. Here, we compare
the slopes of the best-fit gradients ($\beta$ for clumps and $\beta_d$ for
diffuse background in the figure). At \mstar$<10^{10.6}$\msun\ and $z<2.0$,
where we find obvious and strong clump age gradients, the clump age gradient is
significantly steeper than the diffuse background's age gradient. In fact, the
diffuse background only shows an obvious age gradient for low-mass galaxies
with \mstar$<10^{9.8}$\msun. Moreover, at $z<2.0$, the age of inner clumps
($d_{clump}/SMA<0.5$) is older than that of the inner part of the diffuse
background.

In our SED-fitting, the lower limit of the age of stellar population models is
10 Myr (see Table \ref{tb:sedparam}). Using younger models in our SED-fitting
procedure would result in an unphysically high SFR for some clumps. As a result
of the age limit, a small fraction of clumps are stalled at $log(age/Gyr) \sim
-2$ (more obviously seen in the two panels of $0.5\leq z < 1.0$ and
\mstar$<10^{10.6}$\msun). For these clumps, we likely overestimate their ages.
This caveat, however, does not significantly bias our age gradient
measurement, because the number of these possibly very young clumps is small.
In observation, so far only one clump \citep{zanella15} was measured to be
younger than 10 Myr. Although some simulations, e.g., FIRE, predict very
short-lived clumps, our lower age limit is still younger than the mean lifetime
of their massive clumps ($\sim$20 Myr, \citet{oklopcic17}), enabling our
measurements enough diagnostic power to test their models.


\begin{figure*}[htbp] \center{
\includegraphics[scale=0.4, angle=0]{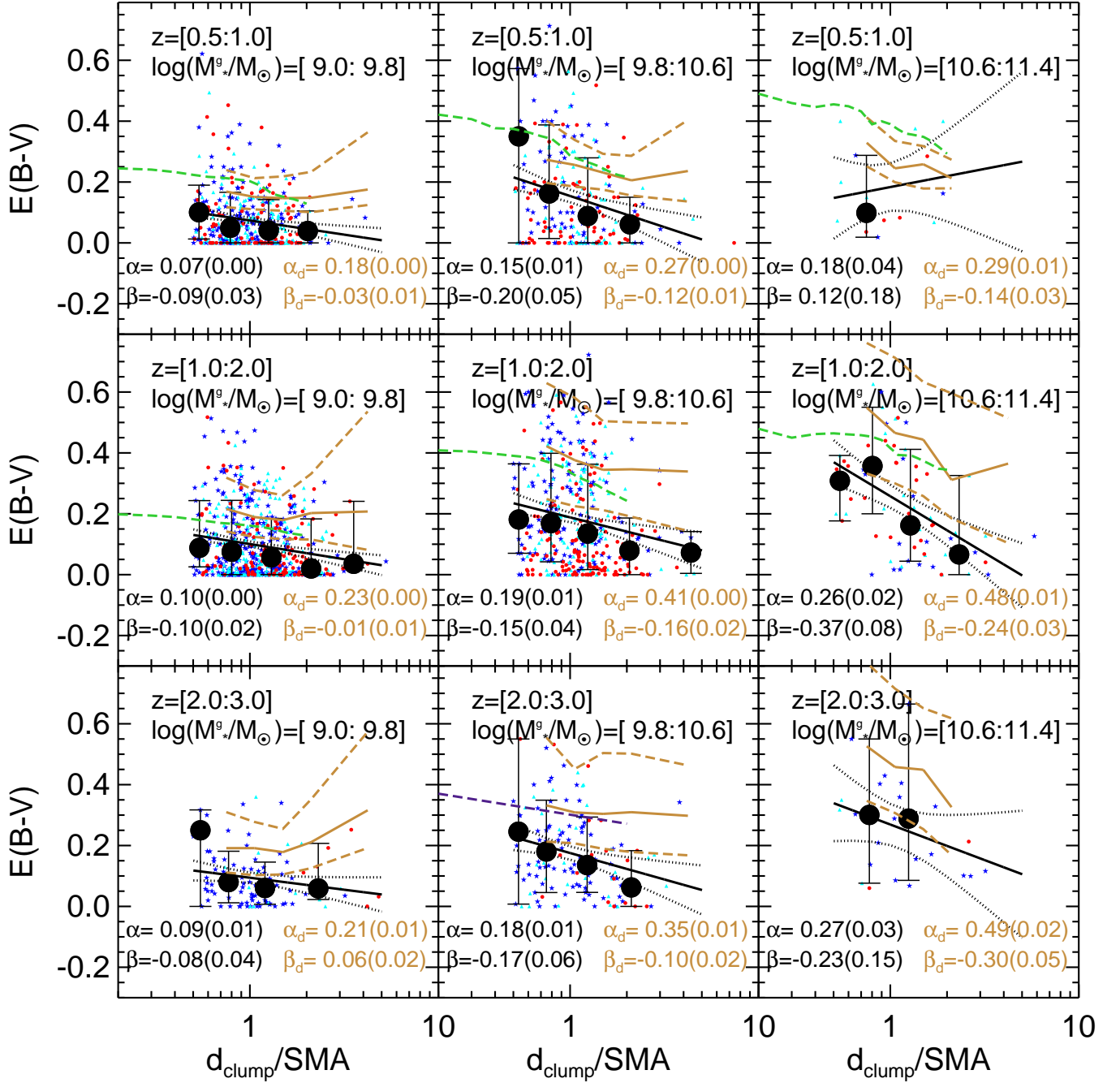}}

\caption[]{Similar to Figure \ref{fig:clgrad_age}, but for the radial gradient
of dust extinction, E(B-V), of clumps. Green and purple dashed lines show the
E(B-V) profiles measured by \citet{wangwc17} and \citet{tacchella17}.

\label{fig:clgrad_ebmv}}
\end{figure*}

\begin{figure*}[htbp] \center{
\includegraphics[scale=0.4, angle=0]{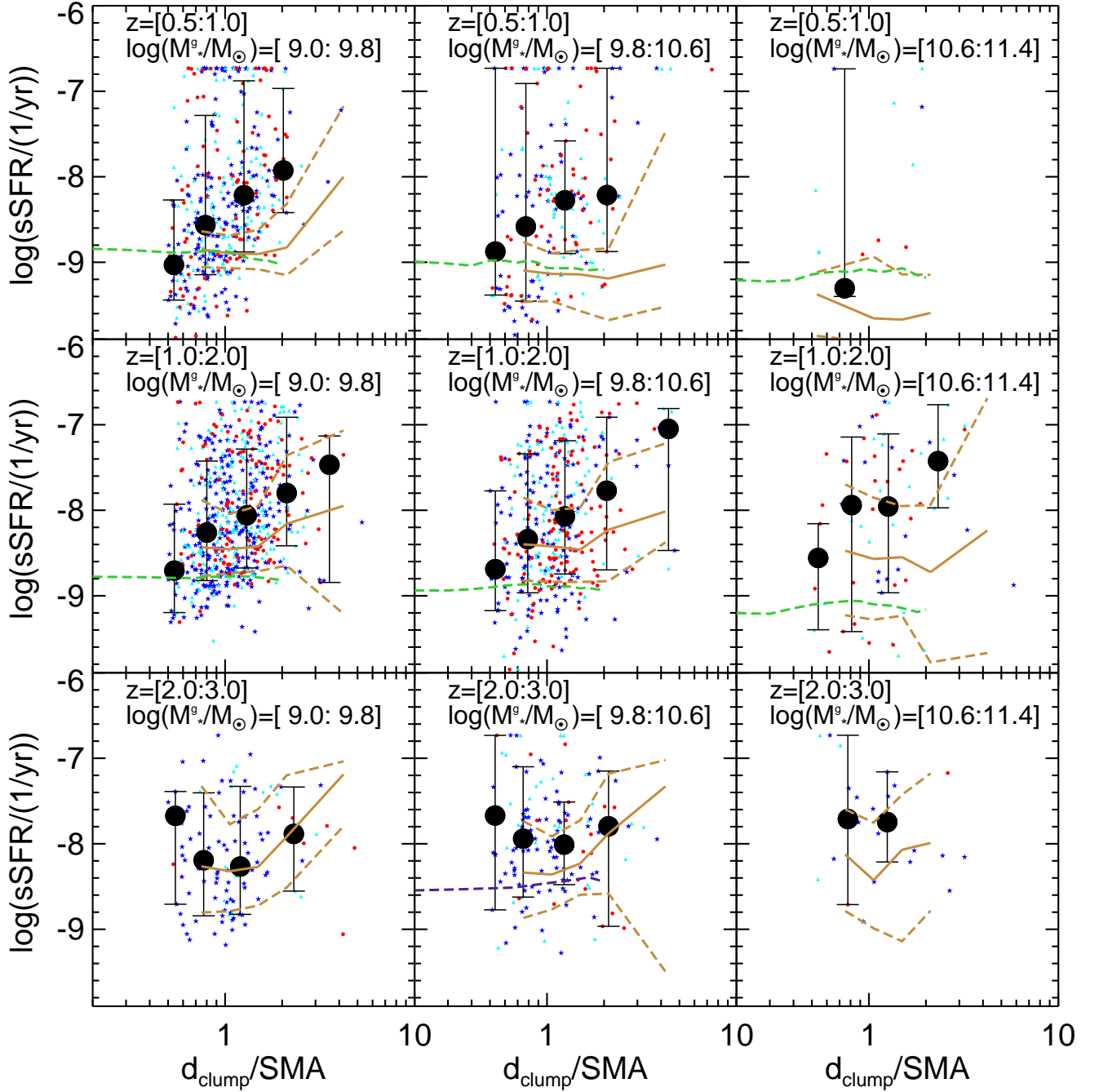}}
\caption[]{Similar to Figure \ref{fig:clgrad_ebmv}, but for the radial gradient
of specific SFR of clumps and diffuse background.
\label{fig:clgrad_ssfr}}
\end{figure*}

\begin{figure*}[htbp] \center{
\includegraphics[scale=0.4, angle=0]{fig12.ps}}

\caption[]{Similar to Figure \ref{fig:clgrad_age}, but for the radial gradient
of \mstar\ of clumps. 

\label{fig:clgrad_mstar}}
\end{figure*}

\begin{figure*}[htbp] \center{
\includegraphics[scale=0.4, angle=0]{fig13.ps}}

\caption[]{Similar to Figure \ref{fig:clgrad_age}, but for the radial gradient
of \mstar\ surface density of clumps. Magenta dashed lines and pink dotted
lines show the profiles measured by \citet{nelson16b} and \citet{mosleh17},
respectively.

\label{fig:clgrad_sigmass}}
\end{figure*}

\subsection{Dust Extinction Gradient}
\label{prop:clgrad_ebmv}

Figure \ref{fig:clgrad_ebmv} shows the radial variation of clump dust
extinction, E(B-V), as a function of galactocentric distance. Overall, clumps
show a negative E(B-V) gradient: inner clumps are more dust-extincted, while
outer clumps have little extinction. The slopes of the clump E(B-V) gradients
depend on galaxy \mstar: at a given redshift, the gradient becomes steeper as
galaxy \mstar\ increases. For a given galaxy \mstar, however, the gradient
shows no obvious dependence on redshift.

Because of the age--dust degeneracy, the observed clump U-V color gradient
(Figure \ref{fig:clgrad_color}) can be explained by an age gradient or an
extinction gradient or a combination of both. Our SED-fitting measures both age
and E(B-V) simultaneously and attributes the observed U-V gradient to both age
(Figure \ref{fig:clgrad_age}) and E(B-V) (Figure \ref{fig:clgrad_ebmv}). If
there was no age gradient as shown in Figure \ref{fig:clgrad_age}, the E(B-V)
gradient would be much stronger, and vice versa. 

One way to test our SED-fitting procedure is to study the E(B-V) gradient of
the diffuse component and compare it with the literature. Diffuse background
and clumps have similar slopes for the E(B-V) gradients, but clumps are
systematically less dust-extincted. This result is not surprising, because
clumps are selected as UV-bright regions and hence likely to have less dust
extinction.

We compare our measurements (brown lines in Figure \ref{fig:clgrad_ebmv}) with
those of \citet{wangwc17} (green dashed lines) and \citet{tacchella17} (purple
dashed line). The methods used by Wang et al. and Tacchella et al. are
different from ours. Wang et al.  calibrated the relation between E(B-V) (and
sSFR) and colors in the rest-frame UVI diagram (a substitute of the UVJ
diagram) of integrated galaxies and applied the calibrations to
multi-wavelength multi-aperture photometry.  Tacchella et al. used the
rest-frame UV continuum to measure E(B-V) and sSFR for massive galaxies at
$z\sim2.2$.

The measurements of the two groups, although derived by different methods, show
good agreement with our results. The difference between their and our E(B-V)
profiles is within the scatter of our measurements, except for the most massive
bin at $z<1.0$, where both our and Wang's samples suffer from small number
statistics. Both Wang et al. and Tacchella et al. used all pixels to measure
the profiles, while we only use intra-clump pixels (i.e., those not masked as
clump locations -- see the white pixels in the bottom right panel of Figure
\ref{fig:bkg}). This difference may change the slopes of the E(B-V) profiles in
our and their studies. In this test, however, we only focus on the absolute
values within the galactocentric ranges covered by both our and other studies. 

Overall, this result demonstrates that our SED-fitting procedure induces no
significant systematics compared with other studies. When breaking the
age--dust degeneracy, our method yields consistent results with similar studies
in the literature. Based on our results, we argue that to explain the observed
clump U-V gradient requires both an age gradient and an E(B-V) gradient
simultaneously. \citet{liufs16} also showed that the observed color gradient of
galaxies is composite with both stellar population and E(B-V) gradients. 


An additional method to assess the simultaneous need of both age and dust
extinction gradient is to ``marginalize'' over either age or E(B-V). In this
method, we assume no radial gradient for one quantity and using the other one
to explain the observed UV colors gradients of clumps. This condition
requires clumps near galactic centers to be unrealistically old/dusty. 

We use clumps in galaxies with $9.8 \leq \logm < 10.6$ at $1.0 \leq z < 2.0$ as
an example. The UV colors of these clumps drop from U-V$\sim$0.8 at near
galactic centers to $\sim$0 at $\sim$3$\times$SMA. If we fix E(B-V)=0.0 for all
radii and assume a constant SFH, the inner clumps need to be as old as $\sim$5
Gyr to reach U-V$\sim$0.8, which is older than the age of the universe at
$z\sim1.5$. If we choose a tau-model with $\tau$=0.5 Gyr for the SFH, the inner
clumps would have an age of 1 Gyr.  This age is younger than the age of the
universe, but still two times older than the characteristic timescale of SF,
suggesting the clump SF is being quenched, which is inconsistent with the
prominent UV luminosity of the clumps.

On the other hand, if we fix clump age as 30 Myrs across all radii, the inner
clumps need to have E(B-V)$\sim$0.5 to reach U-V$\sim$0.8. According to the
Calzetti extinction law adapted in our paper, $A_{NUV (2800A)}$ of the
inner clumps would be 3.6 mag, resulting in the dust-corrected NUV luminosity
of inner clumps being $\sim$30 times brighter than what we observed. With such
high attenuation, even a single clump (considering each contributing 5\% of the
``observed'' UV luminosity of the galaxy) would easily have a SFR larger than
the global SFR of the total galaxy. Therefore, we believe that dust extinction
alone cannot fully explain the observed clump U-V gradient. This test of E(B-V)
has little dependence on the choice of clump age and SFH.

\subsection{sSFR Gradient}
\label{prop:clgrad_ssfr}

Figure \ref{fig:clgrad_ssfr} shows the sSFR gradient of clumps. At $z<2.0$
(except for the most massive galaxies,
$10^{10.6}$\msun$<$\mstar$<10^{11.4}$\msun, at $0.5\leq z<1.0$ where the sample
size is tiny), clumps exhibit strong radial variation: sSFR of inner clumps
(at $d_{clump}/SMA=0.5$) is about one dex lower than that of outer clumps
(at $d_{clump}/SMA \gtrsim 0.5$).

Similar to the test of the dust extinction gradient, we also measure the sSFR
gradients of diffuse background (brown lines in the figure) and compare our results with
those of \citet{wangwc17} (green dashed lines) and \citet{tacchella17} (purple
dashed line). Our results show excellent agreement with these studies at
$z<1.0$ and $z\geq2.0$. At at $1.0\leq z < 2.0$, however, our measurements are
higher, although still within the uncertainties, than those of Wang et al. A
possible reason is the redshift distributions: the sample of Wang et al. is in
fact at $z<1.4$, while our sample covers the whole redshift range of $1.0\leq z
< 2.0$. The lack of $z>1.4$ (and more actively star forming) galaxies in Wang
et al. therefore biases their measurement to lower values compared to our
sample. Overall, good agreement between our and other studies ensures the
accuracy of our sSFR gradient measurement.

Our results, together with those of Wang et al. and Tacchella et al., show that
the diffuse background  (or integrated) sSFR gradient (or sSFR profile) is
almost flat from their galactic centers to $2 \times SMA$. Beyond $2 \times
SMA$, our results show marginal evidence of an increasing sSFR toward large
galactocentric distance. At $z<2.0$, the clump sSFR gradient is steeper than
that of the diffuse component: inner clumps have similar or even lower sSFR
than the diffuse background, while outer clumps' sSFR is about $\sim 0.5$
dex higher than that of the background.

\subsection{Stellar Mass and Stellar Mass Density Gradients}
\label{prop:clgrad_sigmass}

Figure \ref{fig:clgrad_mstar} shows the radial variation of \mstar\ of clumps.
In all redshift and galaxy \mstar\ bins where the sample size is large enough,
clumps show a significant \mstar\ gradient: 
inner clumps are on average more massive than outer clumps by a factor of
a few tens to hundred.

Figure \ref{fig:clgrad_sigmass} shows the radial variation of the stellar mass
density ($\Sigma_{*}$) of clumps. $\Sigma_{*}$ is calculated within a circle
with radius of 0\farcs18 (3 pixels), namely the aperture size that is used to
measure the clump photometry. 
Clumps show a significant $\Sigma_{*}$ gradient in almost all redshift and
galaxy \mstar\ bins (except for the two most massive bins with small sample
statistics): inner clumps are on average denser than outer clumps by a
factor of $\sim$10--30.

In fact, the $\Sigma_{*}$ measured above should be treated as the lower limit
of the true clump $\Sigma_{*}$. In this paper, we assume clumps are unresolved
sources and use 0\farcs18 as their radius. This choice is comparable to the
FWHM of {\it HST} F160W, below which sources are unresolved. The actual size of
clumps is uncertain, ranging from $\sim$100 pc to $\sim$1 kpc. Our aperture of
0\farcs18 is equal to $\sim$1.5 kpc at $z\sim1$, which is much larger than most
clump size measurements in the literature. Our clump size is therefore an upper
limit of the size of an unresolved source. Accordingly, the measured
$\Sigma_{*}$ is a lower limit. 
Moreover, the $\Sigma_{*}$ gradient will only be true if clump size has no
dependence on their galactocentric distance.

We also show $\Sigma_{*}$ profiles of diffuse component from our sample (brown
lines in the figure) and those from \citet{nelson16b} (magenta dashed lines)
and \citet{mosleh17} (pink dotted lines). \citet{nelson16b} converted the
observed F140W light profile into a mass surface density profile by applying
the integrated \mstar--to--F140W ratio as a constant scale factor at all radii.
\citet{mosleh17} used FAST \citep{kriek09fast} to measure \mstar\ at each
radius. Both \citet{nelson16b} and \citet{mosleh17} provided $\Sigma_{*}$
profiles in units of kpc. We scaled their galactocentric distance by the
average size of galaxies in their samples in each redshift and \mstar\ bin.
Overall, the agreement is good: the difference between our and their studies is
within the confidence level of our measurements. The slopes of the profiles
differ from one study to another. As we discuss in the comparison of E(B-V)
profiles, we only focus on the absolute values rather than the slopes to test
any significant systematics. Given this purpose, the encouraging comparison
results support the accuracy of our measurement. 

On average, clump $\Sigma_{*}$ is a few times lower than that of the diffuse
background at the same galactocentric distance. \citet{wuyts12} studied
$\Sigma_{*}$ of clump pixels in massive galaxies at $0.5\leq z<2.5$ and found
that when clumps are detected from rest-frame U-band, the $\Sigma_{*}$ of clump
pixels is about 10 times smaller than $\Sigma_{*}$ of diffuse background at the
half-light radius determined in U-band. Our results are broadly consistent with
those of \citet{wuyts12}.

The above result is apparently surprising, as one may expect that clumps are on
average denser than the diffuse component, because clumps represent star
forming regions and, given their estimated ages (tens to hundreds Myr), are
expected to be dominated by stars. A few factors are contributing to this
result. First and most importantly, as discussed above, the clump $\Sigma_{*}$
measurement should be treated as a lower limit. Our aperture of 0\farcs18 is
actually an upper limit of clump size. This value is corresponding to about 1.5
kpc at $z\sim1$, which is much larger than most clump size measurements in the
literature. Suppose the intrinsic clump size is 500 pc (some authors even argue
for smaller ones), the clump mass surface density is underestimated by a factor
of nine with our aperture size. Second, our fiducial background subtraction
reduces the clump mass by a factor of two compared to the case of no
subtraction.



\begin{figure}[htbp] \center{
\hspace*{-0.6cm}
\includegraphics[scale=0.16, angle=0]{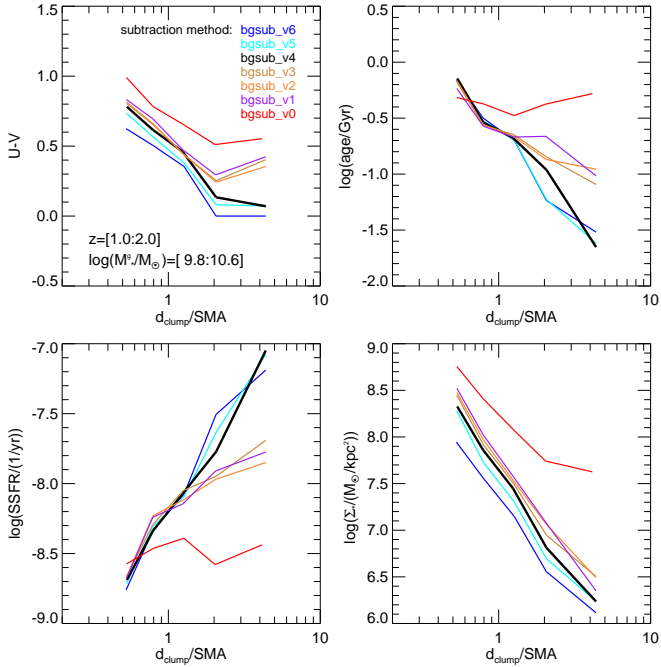}}

\caption[]{Effects of different diffuse background subtractions. As an
example, we show the gradients of of rest-frame U-V color, age, sSFR, and
\mstar\ surface density of clumps in galaxies with $1.0<z<2.0$ and
$10^{9.8}$\msun$<$\mstar$<$$10^{10.6}$\msun. Different color lines show the
median relations with different diffuse background subtraction methods as
indicated by the labels (see Table \ref{tb:bkg} for details). Our fiducial
subtraction method (bgsub\_v4) is shown by the black lines.

\label{fig:aper}}
\end{figure}

\subsection{Effects of Diffuse Background Subtraction}
\label{prop:bkg}

All gradients present in this section are based on our fiducial diffuse
background subtraction (bgsub\_v4 in Table \ref{tb:bkg} and Section
\ref{photometry}). Different subtractions would result in different gradients.
It is important to test if the gradients observed in this section are robust to
different subtraction methods. Particularly, it is interesting to test if
changing the background subtraction would make the age gradient flat.  As
argued by \citet{buck16} and \citet{oklopcic17}, the negative clump age
gradient (inner old, outer young) may be a result of clumps being
contaminated by disk stars. To test the robustness, we repeat all previous
measurements with different background subtraction methods in Table
\ref{tb:bkg}. Here, we only use one redshift and galaxy \mstar\ bin:
$1.0<z<2.0$ and $10^{9.8}$\msun$<$\mstar$<$$10^{10.6}$\msun\ to illustrate the
effects. This bin is representative and may be of interest to many readers. 

The test results are shown in Figure \ref{fig:aper}. For the U-V color, the
overall trend of color gradient is preserved from very aggressive subtraction
(bgsub\_v6, blue in the figure) to no subtraction (bgsub\_v0, red), although
the clump colors become redder gradually from bgsub\_v6 to bgsub\_v0. This is
expected because background stars are older than clump stars (Figure
\ref{fig:clgrad_color}), and adding diffuse background light to clumps makes the latter
redder. Moreover, the panel shows that outer clumps are affected by
background subtraction more than inner clumps are. 

Age and sSFR gradients are significantly affected by background subtraction.
From very aggressive subtraction (bgsub\_v6, blue) to very conservative
subtraction (bgsub\_v1, purple), both gradients become flatter. Eventually,
when no background subtraction is applied (bgsub\_v0, red), both gradients
become flat. This change is again mostly driven by the effects of outer
clumps. From bgsub\_v6 to bgsub\_v1, the age (sSFR) of clumps at $4 \times
d_{clump}/SMA$ becomes older (smaller) by a factor of three (six). Inner
clumps, however, are hardly changed in various subtractions. For example,
in most of the subtraction methods, the sSFR and age of the clumps with
galactocentric distance smaller than SMA are little changed.

\mstar\ surface density is similar to the U-V color: the overall trend is
preserved, but the amplitude changes. From bgsub\_v1 (purple) to bgsub\_v6
(blue), $\Sigma_{*}$ decreases by a factor of three, because more and more
light is subtracted from clumps. No subtraction (bgsub\_v0, red) is even denser
than the aggressive subtraction (bgsub\_v6, blue) by another factor of three,
approaching the $\Sigma_{*}$ profile of diffuse background (light brown,
magenta, and pink lines in the middle panel of Figure
\ref{fig:clgrad_sigmass}).

Overall, we conclude that background subtraction methods (except for the no
subtraction one) would not change our conclusions of the existence of the U-V,
age, sSFR, and $\Sigma_{*}$ gradients of clumps, although they may alter the
amplitude and slope of each gradient. Also, outer clumps are more
vulnerable to background subtraction than inner clumps are.

\section{Catalog Release and Use}
\label{catalog}

We release the clump catalogs with the electronic version of this paper. The
columns of the catalog are described in Table \ref{tb:cat}. We release the
clump parameters measured with all background subtraction methods in Table
\ref{tb:bkg}. In the released catalog, Columns 19-68 are repeated for each
background subtraction method (the method is given by Column 18).

We also suggest readers to apply the following criteria to exclude clumps
with problematic photometry or derived properties: 
\begin{enumerate}

\item {\it badczflag}=1: these clumps have catastrophic photo-zs compared with
the redshifts of their galaxies, which implies photometric errors (see Section
\ref{photometry:accuracy}).

\item {\it veryredflag}=1: these clumps have rest-frame U-V colors redder than
the separation of blue cloud and red sequence of integrated galaxies, which
also indicates a photometry problem (see Section \ref{measureprop:cm}).

\item {\it badmassflag}=1: these clumps' \mstar\ is larger than that of their
host galaxies, indicating an error in either photometry or SED-fitting (see
Section \ref{measureprop:global}).

\item {\it extremesfrflg}=1: these clumps' log(SFR/(\msun/yr)) is larger than 3
or smaller than -2, indicating that the SED-fitting chooses an extreme
solution, possibly due to an error in either photometry or SED-fitting.

\end{enumerate}

\begin{table*}[htbp]
\caption{Clump Catalog Columns \label{tb:cat}}
\begin{tabular}{cccc}
\hline\hline
Column & Name &  Note & Reference \\
\hline
Part I: Galaxy properties \\
\hline
1 & Galaxy ID & CANDELS ID & \citet{ycguo13goodss} \\
2 & RA & J2000 &  --- \\
3 & DEC & J2000 &  --- \\
4 & Redshift & & \citet{dahlen13} \\
5 & \mstar\ &  ${\rm log}$\msun\ & \citet{mobasher15,santini15} \\
6 & \mstar\ error &  dex & --- \\
7 & SFR &  ${\rm log}$(\msun/yr) & --- \\
8 & SFR error &  dex & --- \\
9 & Rest-frame U & magnitude & \\
10 & Rest-frame V & magnitude & \\
11 & Rest-frame J & magnitude & \\
12 & SMA & arcsec & \citet{vanderwel14size} \\
13 & SMA error & arcsec & --- \\
\hline
Part II: Observed Clump properties \\
\hline
14 & Clump ID & & \citet{ycguo15fclumpy} \\
15 & Clump RA & J2000 & --- \\
16 & Clump DEC & J2000 & --- \\
17 & Clump detection band & & \citet{ycguo15fclumpy} or Section \ref{sample:clump} \\
18 & Clump $f_{LUV}$ & & \citet{ycguo15fclumpy} \\
19 & Background subtraction configuration & & Section \ref{photometry:bkg} \& Table \ref{tb:bkg} \\
20,21 & F435W flux and error & $\mu$Jy & Section \ref{photometry} \\
22,23 & F606W flux and error & $\mu$Jy & --- \\
24,25 & F775W flux and error & $\mu$Jy & --- \\
26,27 & F814W flux and error & $\mu$Jy & --- \\
28,29 & F850LP flux and error & $\mu$Jy & --- \\
30,31 & F105W flux and error & $\mu$Jy & --- \\
32,33 & F125W flux and error & $\mu$Jy & --- \\
34,35 & F140W flux and error & $\mu$Jy & --- \\
36,37 & F160W flux and error & $\mu$Jy & --- \\
38 & Galactocentric distance & normalized by galaxy SMA & \citet{ycguo15fclumpy} \\
\hline
Part III: Derived Clump properties \\
\hline
39 & Clump photo-z & & Section \ref{photometry:accuracy} \\
40 & {\it badczflag} & & Section \ref{photometry:accuracy} \\
41,42,43 & \mstar\ and its lower and upper 1$\sigma$ & ${\rm log}$\msun\ & Section \ref{measureprop:sed} \\
44,45,46 & SFR and its lower and upper 1$\sigma$ & ${\rm log}$(\msun/yr) & --- \\
47,48,49 & E(B-V) and its lower and upper 1$\sigma$ & & --- \\
50,51,52 & Age and its lower and upper 1$\sigma$ & ${\rm log}$(Gyr) & --- \\
53,54,55 & $\tau$ and its lower and upper 1$\sigma$ & ${\rm log}$(Gyr) & --- \\
56,57 & Rest-frame U and error & magnitude & --- \\
58,59 & Rest-frame B and error & magnitude & --- \\
60,61 & Rest-frame V and error & magnitude & --- \\
62,63 & UV SFR and error & ${\rm log}$(\msun/yr), dex & Section \ref{measureprop:sfr} \\
64,65 & UV E(B-V) and error & & --- \\
66 & {\it veryredflag} & & Section \ref{measureprop:cm} \\
67 & {\it badmassflag} & & Section \ref{catalog} \\
68 & {\it extremesfrflag} & & --- \\
\hline
\end{tabular}
\end{table*}


\section{Summary}
\label{summmary}

As a step to establish a benchmark of direct comparisons of clumps between
observations and theoretical models, we present a sample of clumps, which, to
the best of our knowledge, represents the commonly observed non-lensed
``clumps'' discussed in the literature. This sample contains 3193 clumps
detected from 1270 galaxies at $0.5 \leq z < 3.0$. The clumps are detected
from rest-frame UV images as described in Paper I. The physical properties of
the clumps, e.g., rest-frame color, \mstar, SFR, age, and dust extinction, are
measured through fitting clump SEDs to synthetic stellar population models. 

We carefully test the procedures of measuring clump properties in a few ways:
(1) clump mass--to--light ratio, (2) clump color--\mstar\ diagram, (3) SFRs and
E(B-V)s measured by both SED-fitting and UV continuum, and (4) as an indirect
test, comparisons between our and others' measurements of the radial profiles
of physical properties. We also test the effects of subtracting background
fluxes from the diffuse component (or intra-clump regions) 
of galaxies on the observed clump gradients.

We show some examples of the measured physical properties. 
We find clumps show radial U-V color variation: clumps close to galactic
centers are redder than those in outskirts. The slope of the color gradient
(clump color as a function of their galactocentric distance scaled by the
semi-major axis of galaxies) changes with redshift and stellar mass of the host
galaxies: at a fixed stellar mass, it becomes steeper toward low redshift; and
at a fixed redshift, it becomes steeper toward massive galaxies. Based on our
SED-fitting, this observed color gradient can be explained by a combination of
a negative age gradient, a negative E(B-V) gradient, and a positive specific
star formation rate gradient of clumps. The color gradients of clumps are
steeper than those of intra-clump regions (``disks''). Correspondingly, The
radial gradients of the derived physical properties of clumps are different
from those of the diffuse component (intra-clump regions or ``disks'').

\ \ \ 

\ \ \

We would like to thank the anonymous referee for carefully reviewing our
paper and providing constructive comments, which truly improved the quality of
the paper. Support for Program HST-GO-12060 and HST-GO-13309 was provided by
NASA through a grant from the Space Telescope Science Institute, which is
operated by the Association of Universities for Research in Astronomy,
Incorporated, under NASA contract NAS5-26555. P.G.P.-G. acknowledges support
from Spanish Government MINECO AYA2015-70815-ERC and AYA2015-63650-P Grants. 

\

{\it Facilities}: \hst\ (ACS and WFC3)



\bibliographystyle{apj}
\bibliography{references}

\end{document}